%% file: 3g_cooling.tex
\documentclass[
  aps,
  prd,
  reprint,
  nofootinbib,
  noeprint,
]{revtex4-2}

\input{header}

\input{definitions}

\usetikzlibrary{external}
\begin{document}

\title{True and apparent motion of optomechanical resonators, with applications to feedback cooling of gravitational wave detector test masses}
\author{Evan D. Hall}
\author{Kevin Kuns}
\affiliation{LIGO Laboratory, Department of Physics, Massachusetts Institute of Technology, 185 Albany St, Cambridge, MA 02139, USA}

\begin{abstract}
  Modern optomechanical systems employ increasingly sophisticated quantum-mechanical states of light to probe and manipulate mechanical motion.
  Squeezed states are now used routinely to enhance the sensitivity of gravitational-wave interferometers to small external forces, and they are also used in feedback-based trapping and damping experiments on the same interferometers to enhance the achievable cooling of fluctuations in the differential test mass mode~\cite{Whittle:2021mtt}.
  In this latter context, an accurate accounting of the true test mass motion, incorporating all sources of loss, the effect of feedback control, and the influence of classical force and sensing noises, is paramount.
  We work within the two-photon formalism to provide such an accounting, which extends a previously described decomposition of the quantum-mechanical noise of the light field~\cite{McCuller:2021mbn}.
  This decomposition provides insight, rooted in physically motivated parameters, into the optimal squeezed state and feedback control configuration that should be employed to achieve the lowest fluctuations.
  We apply this formalism to feedback damping experiments in current and possible future gravitational-wave interferometers\,---\,LIGO A+, LIGO Voyager, Cosmic Explorer (CE), and CE Voyager\,---\,and discuss how these multi-degree-of-freedom systems might be compared to a single degree-of-freedom oscillator.
  We find that, for the oscillator definition used most commonly in the literature so far, occupation numbers below 1 are possible in these interferometers over a frequency range comparable to the bandwidth of the trapped and cooled oscillator.
  We also discuss several technical issues in cooling experiments with gravitational-wave detectors.
\end{abstract}

\maketitle

\input{sections/introduction}

\input{sections/perspective}

\input{sections/true_motion}

\input{sections/thermometry}
\input{sections/ligo_ce}

\input{sections/technical}

\input{sections/conclusion}

\begin{acknowledgments}
The authors thank Lee McCuller for insightful discussions and for the initial implementation of the software and display framework utilized for the signal flow graph algebra.
EDH and KK are supported by NSF PHY--2309200 and PHY--2309064. KK is additionally supported by NSF PHY--2309267, and EDH is additionally supported by PHY--2308972.
This material is based upon work supported by NSF's LIGO Laboratory, which is a major facility fully funded by the US National Science Foundation.
LIGO was constructed by the California Institute of Technology and Massachusetts Institute of Technology with funding from the NSF and operates under NSF Cooperative Agreement PHY--2309200. Advanced LIGO was built under NSF PHY--0823459.
The authors also gratefully acknowledge the support of the Science and Technology Facilities Council (STFC) of the United Kingdom, the Max-Planck-Society (MPS), and the State of Niedersachsen/Germany for support of the construction of Advanced LIGO and construction and operation of the GEO\,600 detector; additional support for Advanced LIGO was provided by the Australian Research Council.
The LIGO~A+ Upgrade to Advanced LIGO is supported by NSF PHY--1834382 and STFC ST/S00246/1, with additional support from the Australian Research Council.
\end{acknowledgments}

\appendix
\input{sections/radiation_pressure_mirror}
\input{sections/ff_app}

\bibliography{3g_cooling.bib}

\end{document}

%% file: header.tex
\usepackage[english]{babel}
\usepackage{amsfonts}%
\usepackage{mathtools}
\usepackage{microtype}
\usepackage{verbatim}
\usepackage[mathcal]{euscript}
\usepackage[bb=boondox]{mathalpha}

\usepackage[
  cmyk,
  usenames,
  svgnames,
  dvipsnames,
]{xcolor}
\usepackage{graphicx}
\graphicspath{{./figures/}}
\usepackage{tikz}
\usetikzlibrary{decorations.markings,decorations.pathreplacing,arrows.meta,bending}

\usepackage[
  colorlinks=true,
  citecolor=MediumBlue,
  linkcolor=MediumBlue,
  urlcolor=DarkGreen,
]{hyperref}

\usepackage[capitalise]{cleveref}
\usepackage{enumitem}
\setlist[enumerate]{leftmargin=1ex}
\usepackage{siunitx}
\usepackage{natbib}

%% file: definitions.tex
\tikzset{
  semithick,
  midarrow/.style={decoration={markings, mark=at position 0.5 with {\arrow{Stealth[round,scale=1.35,flex,sep=-3pt]}}},postaction={decorate}},
  dot/.style = {circle, fill, minimum size=#1, inner sep=0pt, outer sep=0pt},
  dot/.default = 4pt  %
}

\newcommand{\rmd}{\mathrm{d}}
\newcommand{\rme}{\mathrm{e}}
\newcommand{\rmi}{\mathrm{i}}

\newcommand{\mathbbl}{\mathbb}

\newcommand{\Tmat}[1]{\ensuremath{\mathbb{#1}}}
\newcommand{\Tvec}[1]{\vec{\mathbb{#1}}}

\newcommand{\Krp}{\ensuremath{\mathcal{K}}}

\newcommand{\Gctrl}{\ensuremath{G_\text{ctrl}}}
\newcommand{\Fext}{\ensuremath{F_\text{ext}}}
\newcommand{\xsens}{\ensuremath{x_\text{sens}}}
\newcommand{\om}[1]{#1_\text{om}}
\newcommand{\eff}[1]{#1_\text{eff}}
\newcommand{\mrse}[1]{#1_\text{rse}}
\newcommand{\osc}[1]{#1_\text{osc}}
\newcommand{\sql}{\ensuremath{\Omega_{\text{sql}}}}
\newcommand{\zpf}{\ensuremath{x_{\text{zpf}}}}
\newcommand{\rrse}{\ensuremath{\mathfrak{r}_{\text{rse}}}}
\newcommand{\trse}{\ensuremath{\mathfrak{t}_{\text{rse}}}}
\newcommand{\Fa}{\mathcal{F}_{\text{a}}}
\newcommand{\Fs}{\mathcal{F}_{\text{s}}}

\newcommand{\aux}[1]{#1_{\text{aux}}}
\newcommand{\kappad}{\kappa_{\text{d}}}
\newcommand{\kappas}{\kappa_{\text{s}}}

\DeclareMathOperator{\imag}{Im}
\DeclareMathOperator{\real}{Re}

%% file: sections/introduction.tex
\section{Introduction}
\label{sec:introduction}

The basic picture of quantum fluctuations in gravitational-wave interferometers, and similar optomechanical devices, has been known for more than forty years~\cite{Caves:1981hw}.
In this picture, quantum fluctuations of the two quadratures of the optical field entering the interferometer's readout port give rise to phase noise and test mass radiation pressure noise, thereby sourcing fluctuations on the two quadratures of the optical field exiting the interferometer's readout port.
This picture has proven widely applicable, facilitating a quantum-mechanical description of a variety of optomechanical phenomena.
In particular, it has provided a quantum-mechanical description of the feedback damping of harmonic oscillators~\cite{Mancini:1998wf,Vitali:2001dp,2001EPJD...17..399C,2003JOSAB..20.1054V,Genes:2007bw}.
Initiated with several electromechanical experiments~\cite{Milatz:1953,Roll:1964rd,KittelStatistical}, feedback damping can reduce the system's motional fluctuations far below the level of thermally driven motion that would be present in the absence of feedback.
Based on this residual fluctuation, it is possible to assign an effective temperature, and hence an effective phonon occupation number, to a harmonic oscillator system that has been cooled by feedback damping.
This has now been applied to systems with masses across more than 30 orders of magnitude, with several experiments at the upper end of this range having been performed on gravitational-wave detectors, specifically in the resonant bar detector AURIGA (4000 phonons in a \qty{1100}{\kg} mode)~\cite{Vinante:2008zqr}, as well as laser interferometers Initial LIGO (200 phonons in a \qty{2.7}{\kg} mode)~\cite{LIGOScientific:2009mif} and Advanced LIGO (11 phonons in a \qty{10}{\kg} mode)~\cite{Whittle:2021mtt}.

Feedback cooling does not, in general, enhance the sensitivity of optomechanical force sensors like gravitational-wave detectors.
Rather, it is typically invoked as a means to prepare a harmonic oscillator in a manifestly quantum-mechanical state, such as the ground state or a superposition of two states (e.g.,~\cite{Marshall:2002exi,Romero-Isart:2011sdw}).
To give one example of an application in fundamental physics, a number of authors have proposed to witness the presence (or absence) of gravitationally-mediated entanglement in the motion of harmonic oscillators~\cite{Carney:2018ofe,Krisnanda:2019glc,Miao:2019pxw,Datta:2021ywm}, which would constitute evidence for a quantum theory of gravity~\cite{Bose:2017nin,Marletto:2017kzi}.

Particularly for gravitational-wave interferometers, the full picture of quantum noise is more complicated than the single-mode description: quantum noise is additionally sourced by optical loss~\cite{Kimble:2000gu,Miao:2018pai}, and is affected by the compound optical cavities used to enhance the stored power and set the instrument bandwidth~\cite{Buonanno:2001cj,Miao:2018pai}.
The story has grown even more complicated as these interferometers now employ squeezed vacuum states~\cite{Barsotti:2018hvm,Grote:2013oio,Tse:2019wcy,Virgo:2019juy}, along with filter cavities to reduce both sensing noise and radiation pressure noise at the appropriate frequencies to achieve broadband astrophysical sensitivity enhancement~\cite{LIGOO4Detector:2023wmz}; these additions further complicate the analysis of the quantum optical field entering, circulating in, and exiting the interferometer~\cite{McCuller:2021mbn,Harms:2003hn,Barsotti:2018hvm}.

Therefore, while gravitational-wave interferometers may present attractive systems for feedback cooling, a proper analysis requires considerable care.
This work sets out to do just that, using an input--output formalism frequently used to analyze gravitational wave detectors~\cite{Kimble:2000gu,Buonanno:2001cj,Harms:2003hn,Miao:2018pai,McCuller:2021mbn,2012emqm.book.....M,Danilishin:2012fa} extended to include the true motion of the mirrors and the effects of feedback control.

\subsection{Summary of results}

Our main result, derived in \cref{sec:true_motion}, is summarized by our expression for the spectral density of the true motion of an oscillator that has been trapped and cooled by the application of measurement-based feedback control, \cref{eq:general_true_motion_spectrum}.
This is to be contrasted with the expression for the spectrum of the oscillator's apparent motion as measured optically, \cref{eq:general_measurement_spectrum}.
We arrive at these expressions by considering the flow of classical and quantum fluctuations through the optomechanical system (\cref{fig:signal_flow_reduced}), with the ability to accommodate nontrivial Gaussian states such as squeezed light, as well as feedback control.
These expressions properly account for motional and optical correlations that give rise to ``noise squashing'' and other phenomena that complicate the inference of the true motion of the oscillator from the measurement channel.

This formalism also takes advantage of a physically motivated decomposition of the quantum noise, revealing the role of losses, fundamental phase noise, and the rotation of the injected squeezed state as it interacts with the optomechanical system~\cite{McCuller:2021mbn}. In particular, it is shown that

\begin{enumerate}[leftmargin=2\parindent,rightmargin=1\parindent]
\item Injecting a squeezed state into the interferometer can reduce the quantum noise contribution to the true motion; however, since the rotation of the squeezed state responsible for this motion is different from that at the readout, it is not advantageous to use a filter cavity to inject a frequency dependent squeezed state as is done to reduce measurement noise in the routine operation of gravitational wave detectors.
\item The feedback used to cool the oscillator increases the quantum noise by generating a fundamental phase noise which limits the amount of squeezing that can be used. Rather than being caused by optical losses, which is the case for the measurement during the routine operation of gravitational wave detectors, this phase noise is due to the oscillator acquiring a lossy effective susceptibility through the velocity-damped feedback.
\end{enumerate}

After defining an effective mode occupation number for the oscillator motion (\cref{sec:thermometry}), we apply our quantum noise formalism to the case of feedback cooling of the test masses in present and future gravitational-wave detectors LIGO and Cosmic Explorer (\cref{sec:ligo_ce}).
We find that for both of these detectors operating at their design sensitivity, cooling below the ground state is possible.
Certain technical aspects of the sensing and control of the detector\,---\,particularly the decoupling of other optical degrees of freedom from the main differential arm length degree of freedom, and the handling of local gravity fluctuations at the test masses\,---\,need to be reconsidered and perhaps modified from their normal operation in order to achieve the desired cooling; this is described in \cref{sec:technical}.

%% file: sections/perspective.tex
\section{Perspective and overview of methods}
\label{sec:big_picture}

Before delving into the full computation, we shall give a high-level overview of the goals and methods in this work.
The central goal is to arrive at a budget of all the noises making up the true motion (displacement) $x$ of the differential arm length degree of freedom.
This true motion must be distinguished from the measurement (or readout) variable $y$, which is derived from a homodyne photocurrent generated by the light exiting the interferometer's dark port; it is normally this quantity\,---\,which can be calibrated into an apparent displacement\,---\,that is of interest in characterizing the sensitivity of gravitational-wave detectors.
Both quantities $x$ and $y$ contain contributions from external forces $F_{\text{ext}}$.
Additionally, $y$ contains contributions from sensing noise (e.g., phase noise arising from fluctuations of the mirror optical coatings), which can be expressed as an equivalent displacement noise $x_{\text{sens}}$.

\begin{figure*}[t]
    \centering
    \includegraphics[width=0.9\textwidth]{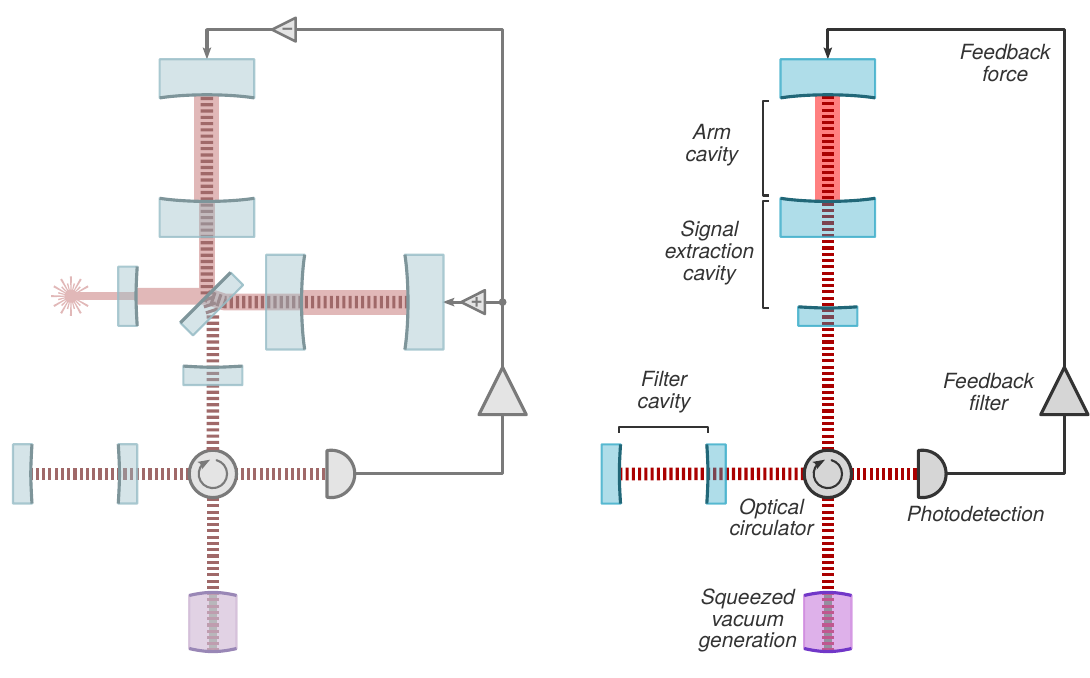}
    \caption{A Fabry--Perot Michelson interferometer with power recycling and signal extraction (\emph{left}) underlying a simplified three-mirror coupled cavity model (\emph{right}) for our study of true and apparent motion under measurement-based feedback control.
      In a Fabry--Perot Michelson interferometer, the two arm cavities are pumped with a coherent optical carrier field (solid lines) from a laser on the bright port side of the beamsplitter.
    Our effective three-mirror description does not include such a port, so we simply assume the presence of an optical carrier field in the arm that does not propagate into the signal extraction cavity or out the dark port.
    Optical sidebands are extracted from the arms, through the partially transmissive signal extraction cavity, and to the photodetector readout at the interferometer's dark port.
    (The readout uses balanced homodyne detection; the coherent local oscillator field for this is not drawn.)
    The modeling of this system and the relevant optical fields, including loss fields not shown in this diagram, is described in the text and shown in the signal flow graph in \cref{fig:signal_flow_full}.
    A squeezed vacuum source and a two-mirror filter cavity are used to injected squeezed vacuum sidebands (dashed lines) into the coupled cavity via an optical circulator.
    The photodetector output is used as an error signal to apply a feedback control force to the end mirror of the arm cavity; in gravitational-wave detectors, this feedback control force is typically applied by electrostatic or magnetic actuation.
    }
    \label{fig:sec_diagram}
\end{figure*}

The introduction of a feedback control force $F_{\text{fb}}$, based on a linear filter of the measurement $y$, alters the dynamics of the system and alters the noise contributions to $x$ and $y$.
In particular, feedback will cause sensing noise $x_{\text{sens}}$ appearing in $y$ to be converted into true test mass motion in $x$.
While the feedback filter can be arbitrary (subject only to stability constraints or technical limitations), for the purposes of feedback cooling the filter can be tailored to produce system dynamics that approximate a damped harmonic oscillator.
If the uncontrolled differential arm length degree of freedom has free-mass dynamics in the frequency range of interest (i.e., the susceptibility is $\chi_0(\Omega) \simeq -1/M\Omega^2$ at each Fourier frequency $\Omega$), then the appropriate feedback control is given by
\begin{equation}
  F_{\text{fb}}(\Omega) = -(M \Omega_{\text{fb}}^2 + \rmi M \Omega \Omega_{\text{fb}} / Q_{\text{fb}} ) \, y(\Omega).
\end{equation}
Here the first term is a feedback-induced spring constant (the trap), and the second term is the feedback-induced damping, with damping rate $\Omega_{\text{fb}} / Q_{\text{fb}}$.

Quantum noise of the optical field can be incorporated into this picture with varying levels of sophistication.
If quantum noise only enters the system at one optical port, then it is sufficient to keep track of the effect of a single pair of ingoing two-photon optical quadratures, which we will denote by the frequency-dependent, two-element vector $\Tvec{s}(\Omega)$.
It is then common to identify one of the two optical quadratures as responsible for radiation-pressure (force) noise $F_{\text{rp}}(\Omega)$, and the other as being responsible for shot noise (a sensing noise) $x_{\text{shot}}(\Omega)$.
However, in more general systems the shot noise and radiation pressure noise may become correlated, rendering this semi-classical treatment insufficient.
Further, when optical losses are present one must consider quantum noise contributions from additional ports, whose quantum noise vectors we collect into the set $\{\Tvec{a}_\mu\}$; in this situation, the total quantum noise similarly does not admit a simple decomposition into radiation-pressure and shot-noise contributions.
In this work we carefully track all the relevant quantum noise vectors, the true differential test mass motion $x$, the measurement $y$, the external forces $F_{\text{ext}}$, the sensing noises, $x_{\text{sens}}$, and all the transfer functions interrelating these quantities.
This is illustrated in \cref{fig:signal_flow_full} for the particular case of a gravitational wave detector.

Among these relations, the optomechanical interaction of light with a moveable mirror is of particular interest (the ``ETM'' subset of the graph shown in \cref{fig:signal_flow_full}).
In the two-photon formalism~\cite{Caves:1985zz,Schumaker:1985zz} that we use, this is represented by a two-element vector transfer function $\Tvec{z}$ that transduces mirror motion $x(\Omega)$ into an outgoing two-photon optical field $\Tvec{a}(\Omega) = \Tvec{z}\,x(\Omega)$.
Conversely, the radiation pressure interaction is represented by a dual vector transfer function $\Tvec{f}^\dagger$ that, when dotted with an ingoing optical field $\Tvec{a}(\Omega)$ at the mirror surface, yields the (scalar) radiation-pressure force $F_{\text{rp}}(\Omega) = \Tvec{f}^\dagger \Tvec{a}(\Omega)$, which induces mirror motion via the mechanical susceptibility $\chi_0(\Omega)$.
These two processes form a radiation-pressure-mediated loop $\chi_0(\Omega)\Tvec{z}\Tvec{f}^\dag$ whereby optical field fluctuations are converted into mechanical motion which is then converted back into optical field fluctuations.
A full accounting of the light-matter interaction in a moveable mirror is given in \cref{sec:mirror_radiation_pressure_dynamics}, where it is shown that this signal flow representation reduces to the familiar input--output relations.

Our formalism applies to any linear optomechanical system in which the mechanical motion can be treated classically, as it can be for the systems discussed in this work (see Ref.~\cite{Whittle:2022gqe} for a discussion of the regimes in which this treatment is valid).
We will ultimately be interested in the dual-recycled Fabry--Perot Michelson topology, which is currently employed for Advanced LIGO~\cite{LIGOScientific:2014pky} and Advanced Virgo~\cite{VIRGO:2014yos}, and is planned for Cosmic Explorer~\cite{Evans:2021gyd} and the Einstein Telescope~\cite{ETDesign2020}.
In this topology (left side of \cref{fig:sec_diagram}), two Fabry--Perot optical cavities, each comprising a suspended input test mass and a suspended end test mass, are arranged with a beamsplitter to form a Michelson interferometer.
One port of the beamsplitter is illuminated with laser light; at the other (dark) port, a photodetector is placed, which registers a photocurrent containing information about the fluctuations of the optical field exiting the antisymmetric port.
A power-recycling mirror is placed between the laser and the beamsplitter to enhance the circulating power in the arm cavities, and a signal extraction mirror is placed between the beamsplitter and the photodetector to adjust the bandwidth of the instrument with respect to differential optical field fluctuations in the arm cavities.
If squeezed vacuum is employed, it is injected into the interferometer from behind the signal-extraction mirror, and the injected squeezed vacuum can be made to have a frequency dependence by the use of an additional optical filter cavity.
For the purposes of modeling the differential motion of the test masses, the dual-recycled Fabry--Perot Michelson topology can be mapped onto a three-mirror compound cavity (right side of \cref{fig:sec_diagram}).
The mathematical aspects of the three-mirror compound cavity are summarized in the signal flow graph in \cref{fig:signal_flow_full}, with the relevant arm cavity radiation pressure dynamics assigned to the end test mass only.
(A full accounting of this graph will be given in \cref{subsec:simplified_ifo}.)

\input{sections/signal_flow_drfpmi}

The following section describes how the quantum-noise fields $\Tvec{s}$ and $\{\Tvec{a}_\mu\}$ and the classical noises $F_\text{ext}$ and $x_\text{sens}$ contribute to the total noise spectra of the true motion $x$ and the measurement $y$; this is presented first with a general formalism, which is then applied to the case of a single mirror (\cref{subsec:single_mirror}) and then to a dual-recycled Fabry--Perot Michelson interferometer of the type used for LIGO and Cosmic Explorer (\cref{subsec:simplified_ifo}).

%% file: sections/signal_flow_drfpmi.tex
\begin{figure*}
  \centering
  \begin{tikzpicture}[
    xshift=1cm,
    y=1cm,
    semithick,
    ]

    \begin{scope}[Dandelion!60!Black,shift={(3,0)}]
      \node[draw,thick,rounded corners,align=center] (itm label) at (0.75,5) {SEM};
      \node[dot] (sem ll) at (0,0) {};
      \node[dot] (sem lr) at (1.5,0) {};
      \node[dot] (sem ul) at (0,3) {};
      \node[dot] (sem ur) at (1.5,3) {};

      \draw[midarrow] (sem lr) -- node[below left] {$-r_\text{s}\mathbbl{1}$} (sem ur);
      \draw[midarrow] (sem ul) -- node[above right] {$r_\text{s}\mathbbl{1}$} (sem ll);
      \draw[midarrow,bend left=10] (sem ul) to node[above] {$t_\text{s}\mathbbl{1}$} (sem ur);
      \draw[midarrow,bend left=10] (sem lr) to node[below] {$t_\text{s}\mathbbl{1}$} (sem ll);
    \end{scope}

    \begin{scope}[Blue,shift={(7.5,-0.5)}]
      \node[draw,thick,rounded corners,align=center] (itm label) at (0.75,5.5) {ITM};
      \node[dot] (itm ll) at (0,0) {};
      \node[dot] (itm lr) at (1.5,0) {};
      \node[dot] (itm ul) at (0,4) {};
      \node[dot] (itm ur) at (1.5,4) {};

      \draw[midarrow] (itm lr) -- node[below left] {$-r_\text{i}\mathbbl{1}$} (itm ur);
      \draw[midarrow] (itm ul) -- node[above right] {$r_\text{i}\mathbbl{1}$} (itm ll);
      \draw[midarrow,bend left=10] (itm ul) to node[above] {$t_\text{i}\mathbbl{1}$} (itm ur);
      \draw[midarrow,bend left=10] (itm lr) to node[below] {$t_\text{i}\mathbbl{1}$} (itm ll);
    \end{scope}

    \begin{scope}[Red!70!Black,shift={(12,-0.5)}]
      \node[draw,thick,rounded corners,align=center] (etm label) at (2,5.5) {ETM};
      \node[dot] (etm ll) at (0,0) {};
      \node[dot] (etm lr) at (4,0) {};
      \node[dot] (etm ul) at (0,4) {};
      \node[dot] (etm ur) at (4,4) {};
      \node[dot] (x1) at (2,2) {};
      \node[dot] (f fi) at (1.15,3.15) {};
      \node[dot] (f fo) at (1.15,0.85) {};
      \node[dot] (f bi) at (2.85,0.85) {};
      \node[dot] (f bo) at (2.85,3.15) {};

      \draw[midarrow] (etm lr) -- node[below right] {$r_\text{e}\mathbbl{1}$} (etm ur);
      \draw[midarrow] (etm ul) -- node[above left] {$-r_\text{e}\mathbbl{1}$} (etm ll);
      \draw[midarrow,bend left=15] (etm ul) to node[above] {$t_\text{e}\mathbbl{1}$} (etm ur);
      \draw[midarrow,bend left=15] (etm lr) to node[below right] {$t_\text{e}\mathbbl{1}$} (etm ll);

      \draw[midarrow,bend right=20] (x1) to node[below right] {$\Tvec{z}_{\text{b}}$} (etm ur);
      \draw[midarrow,bend right=20] (x1) to node[above left] {$\Tvec{z}_{\text{f}}$} (etm ll);

      \draw[midarrow] (etm ul) to node[below] {$\Tvec{f}_{\text{fi}}^\dagger$} (f fi);
      \draw[midarrow] (etm ll) to node[below right] {$\Tvec{f}_{\text{fo}}^\dagger$} (f fo);
      \draw[midarrow] (etm lr) to node[above] {$\Tvec{f}_{\text{bi}}^\dagger$} (f bi);
      \draw[midarrow] (etm ur) to node[above left] {$\Tvec{f}_{\text{bo}}^\dagger$} (f bo);
      \draw[midarrow] (f fi) to node[below left] {$\chi_0$} (x1);
      \draw[midarrow] (f fo) to node[below right] {$\chi_0$} (x1);
      \draw[midarrow] (f bi) to node[above right] {$\chi_0$} (x1);
      \draw[midarrow] (f bo) to node[above left] {$\chi_0$} (x1);
    \end{scope}

    \begin{scope}[Green!70!Black,shift={(3,0)}]
      \node[draw,thick,rounded corners,align=center] (sqz label) at (-1.5,4) {Squeezer};
      \node (a fc) at (0,4.5) {$\Tvec{a}_{\text{fc}}$};
      \node[inner sep=0] (a inj) at (-0.75,1.5) {$\Tvec{a}_{\text{inj}}$};
      \node[dot] (sqz) at (-1.5,3) {};
      \node (svec) at (-2.5,3) {$\Tvec{s}$};

      \draw[midarrow] (svec) -- (sqz);
      \draw[midarrow] (a fc) -- node[above left] {$\Tmat{T}_{\text{fc}}$} (sem ul);
      \draw[midarrow] (a inj) -- node[left] {$\epsilon_{\text{inj}} \mathbbl{1}$} (sem ul);
      \draw[midarrow] (sqz) -- node[above] {$\Tmat{H}_{\text{fc}}$} (sem ul);
    \end{scope}

    \begin{scope}[Plum!80!Black,shift={(3,0)}]
      \node[draw,thick,rounded corners,align=center] (ro label) at (-1.5,0.75) {Readout};
      \node[dot] (y1) at (-1.5,0) {};
      \node (a ro) at (0,-1.25) {$\Tvec{a}_{\text{ro}}$};
      \node (y) at (-2.5,0) {$y$};

      \draw[midarrow] (a ro) -- node[left] {$\epsilon_{\text{ro}} \mathbbl{1}$} (sem ll);
      \draw[midarrow] (sem ll) -- node[above] {$\Tvec{v}^\dagger$} (y1);
      \draw[midarrow] (y1) -- (y);
    \end{scope}

    \begin{scope}[shift={(3,0)}]
      \node (a sec) at (1.5,4.5) {$\Tvec{a}_{\text{sec}}$};

      \draw[midarrow] (a sec) -- node[right] {$\epsilon_{\text{sec}} \mathbbl{1}$} (sem ur);
      \draw[midarrow,out=0,in=180,relative=false] (sem ur) to node[above] {$\Tmat{L}_{\text{s}}$} (itm ul);
      \draw[midarrow,out=180,in=0,relative=false] (itm ll) to node[below] {$\Tmat{L}_{\text{s}}$} (sem lr);
      \draw[midarrow] (itm ur) -- node[above] {$\Tmat{L}_{\text{a}}$} (etm ul);
      \draw[midarrow] (etm ll) -- node[below] {$\Tmat{L}_{\text{a}}$} (itm lr);
    \end{scope}
    
    \begin{scope}[shift={(12,-0.5)}]
      \node (a arm) at (-0.75,1.5) {$\Tvec{a}_{\text{arm}}$};
      \node[inner sep=0] (x sens) at (0,-1) {$x_{\text{sens}}$};
      \node (x) at (5,2.75) {$x$};
      \node (Fext) at (1.5,-1) {$F_{\text{ext}}$};

      \draw[midarrow] (a arm) -- node[left] {$\epsilon_{\text{arm}}\mathbbl{1}$} (etm ll);
      \draw[midarrow] (x sens) -- node[left] {$\Tvec{z}_{\text{f}}$} (etm ll);
      \draw[midarrow,bend right=10] (x1) to (x);
      \draw[midarrow,bend right=20] (Fext) to (f fo);
      \draw[midarrow,out=-75,in=-70] (y1) to node[pos=0.45,above] {$C$} (f fo);
    \end{scope}

    \begin{scope}[Gray!90!Black]
      \draw[decorate,decoration={brace,amplitude=5pt,raise=4ex}]
      (3.05,5) -- (7.45,5) node[midway,yshift=3em]{Signal extraction cavity};
      \draw[decorate,decoration={brace,amplitude=5pt,raise=4ex}]
      (7.55,5) -- (15.95,5) node[midway,yshift=3em]{Arm cavity};
    \end{scope}
    
  \end{tikzpicture}
  \caption{Signal flow graph of a dual-recycled Fabry--Perot Michelson inteferometer with squeezed light injection.
    The system is modeled as a three mirror coupled cavity, with an end test mass (ETM) and input test mass (ITM) comprising a Fabry--Perot cavity, and with a signal extraction mirror (SEM) that broadens the cavity bandwidth. (A diagram is shown in \cref{fig:sec_diagram}.)
    The measurement $y$ is read out in reflection of the interferometer, while the true motion $x$ can only be inferred and cannot be directly measured in an experiment.
    (Freerunning, or apparent, motion $x_\text{free}$ produces the same excitations as $\chi_0\Fext$.)
    The squeezed vacuum state $\Tvec{s}$ is injected into the interferometer after being reflected off of a Fabry--Perot filter cavity.
    For simplicity, the filter cavity has already been algebraically reduced to a reflection matrix $\Tmat{H}_\text{fc}$ and noise transmission matrix $\Tmat{T}_\text{fc}$.
    The measurement $y$ is passed through a control filter $C$ and fed back to the end test mass as a feedback force, which is summed into the same point as $\Fext$.
    Note that quantum radiation pressure and shot noise are sourced by the squeezed field $\Tvec{s}$ and cannot be treated as entering with $\Fext$ or $\xsens$.
    The various sources of unsqueezed vacuum $\{\Tvec{a}_\mu\}$, responsible for the optical losses, are also shown.
  \label{fig:signal_flow_full}}
\end{figure*}
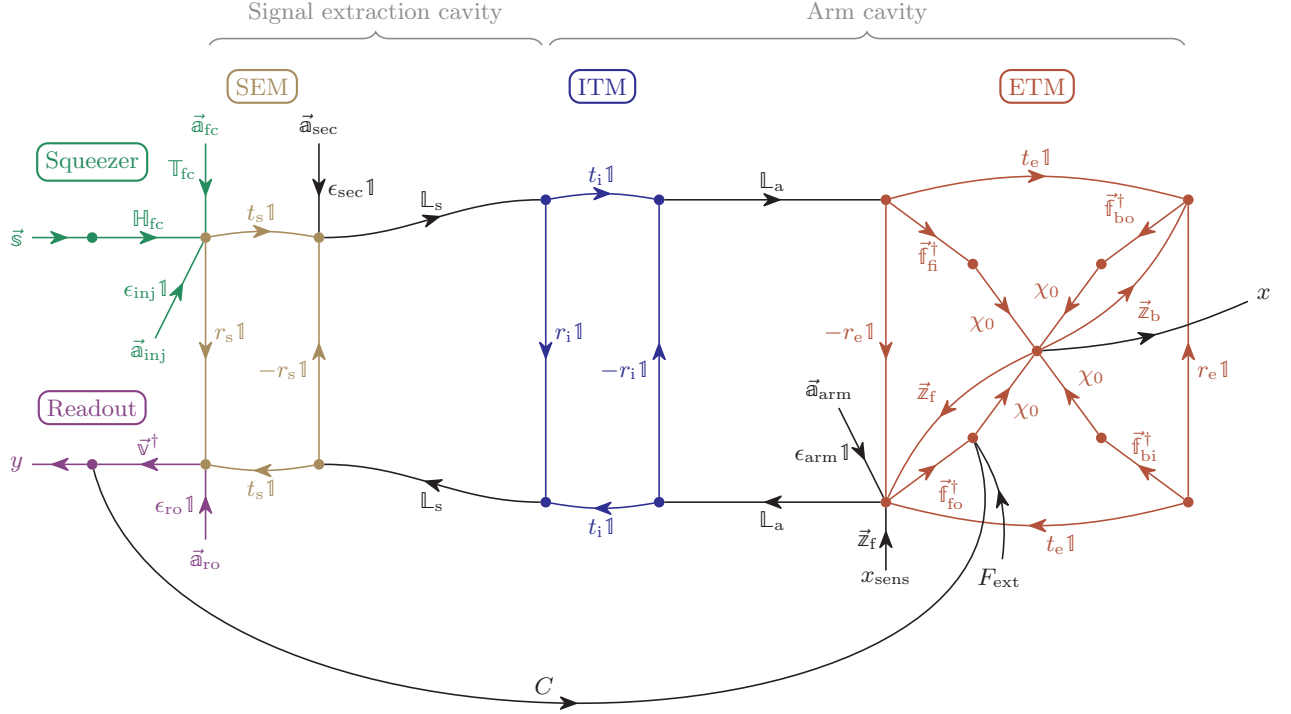

%% file: sections/true_motion.tex
\section{True test mass motion and quantum noise in optomechanical systems}
\label{sec:true_motion}

In \cref{subsec:general_case}, we give a high-level overview of the dynamics of a general system under feedback control. Then in \cref{subsec:single_mirror} we apply these to the simple case of a single mirror under feedback control, and in \cref{subsec:simplified_ifo} describe a simplified system describing gravitational wave interferometers in most cases in more detail.

\input{sections/general_case}

\input{sections/single_mirror}

\input{sections/rse_ifo}

%% file: sections/general_case.tex
\subsection{General case}
\label{subsec:general_case}

The first step in characterizing the true motion $x$ and measurement $y$ is to write down the transfer functions relating these quantities to the optical fields, displacement noises, and external forces that source them, which we do in \cref{subsubsec:general_tfs}.\footnote{In the optical domain specifically, the transfer functions relating field quadratures at various points in the system are referred to as input--output relations.}
We do this with the help of signal flow graphs, which allow us to keep track of these transfer functions, and to algebraically manipulate them in a systematic fashion to provide more tractable expressions.
The dual-recycled Fabry--Perot Michelson interferometer is modeled as a three mirror coupled cavity as shown in \cref{fig:signal_flow_full}.
Once we have these transfer functions, we proceed to compute the spectral densities of $x$ and $y$ in \cref{subsubsec:general_spectra}; here in particular we take advantage of the quantum noise factorization framework developed in Ref.~\cite{McCuller:2021mbn}, which provides physical insight into the origin of the various quantum noise contributions.

\subsubsection{Transfer functions}
\label{subsubsec:general_tfs}

Consider a general optomechanical system in which a mechanical degree of freedom has a true motion $x$, but also in which the optical fields are used to generate a measurement record $y$ that contains information about the same motion. In this work we use the two-photon formalism~\cite{Caves:1985zz,Schumaker:1985zz} where optical fields are described in terms of their amplitude $\hat{q}$ and phase $\hat{p}$ quadratures, and we denote them by vectors
\begin{equation}
  \Tvec{a} =
  \begin{bmatrix}
    \hat{q} \\
    \hat{p}
  \end{bmatrix}
  = \hat{q}\,\Tvec{e}_q + \hat{p}\,\Tvec{e}_p,
  \qquad
  \Tvec{e}_q =
  \begin{bmatrix}
    1 \\
    0
  \end{bmatrix},
  \qquad
  \Tvec{e}_p =
  \begin{bmatrix}
    0 \\
    1
  \end{bmatrix}.
\end{equation}
In the following, the quantities are in general a function of Fourier frequency $\Omega$, but we suppress this argument.
The measurement $y$ itself is obtained by beating the output fields $\Tvec{y}$ with a local oscillator. Mathematically, this local oscillator is denoted by $\Tvec{v}$:
\begin{equation}
  y = \Tvec{v}^\dag \Tvec{y}, \qquad
  \Tvec{v} = \sin\zeta\,\Tvec{e}_q + \cos\zeta\,\Tvec{e}_p,
  \label{eq:homodyne_angle}
\end{equation}
where $\zeta$ is the homodyne angle and $\dag$ denotes Hermitian conjugation.

We will be interested in both external forces $\Fext$ acting on the test mass as well as sensing noises $\xsens$, considering first the system in the absence of feedback control. Furthermore, the quantum state $\Tvec{s}$ entering the system will in general be a squeezed vacuum state. Finally, any sources of loss will couple unsqueezed vacuum $\Tvec{a}_\mu$ into the system. By eliminating all of the optical and mechanical feedback paths, the interferometer as shown in \cref{fig:signal_flow_full} can be reduced and described by several transfer functions as shown in \cref{fig:signal_flow_reduced}. The measurement is
\begin{equation}
  y = \Tvec{v}^\dag \left(
  \om{\Tvec{Z}}\Fext + \om{\Tvec{Y}}\xsens + \om{\Tmat{H}}\Tvec{s}
  + \sum_\mu \Tmat{T}_\mu \Tvec{a}_\mu
  \right),
\end{equation}
where the transfer function $\om{\Tvec{Z}}$ describes how external force on the test mass appears in the output field $\Tvec{y}$, the transfer function $\om{\Tvec{Y}}$ describes how sensing noise appears in $\Tvec{y}$, the transfer function $\om{\Tmat{H}}$ describes how the ingoing squeezed-vacuum quantum noise field appears in $\Tvec{y}$, and the transfer functions $\left\{\Tmat{T}_\mu\right\}$ describe how the unsqueezed vacuum fields $\left\{\Tvec{a}_\mu\right\}$ arising from optical losses appear in $\Tvec{y}$; the index $\mu$ labels the individual loss fields that were specified in \cref{fig:signal_flow_full}.

Conversely, the true motion is
\begin{equation}
  x = \om{\chi}\Fext + \om{X}\xsens + \om{\Tvec{D}}^\dag \Tvec{s}
  + \sum_\mu \Tvec{J}_\mu^\dag \Tvec{a}_\mu,
\end{equation}
which again involves contributions from external force, sensing noise, and the various optical fields, but now with entirely different transfer functions that specify the appearance of these quantities in the motion $x$.\footnote{Rather than the true motion, it is the freerunning, or apparent, motion\,---\,by definition, the motion that would result from the application of an external force $F_{\text{free}} \equiv y/(\Tvec{v}^\dagger \om{\Tvec{Z}})$ via the bare mechanical susceptibility $\chi_0$\,---\,that can be inferred from the measurement $y$:
\begin{equation}
  x_\text{free} \equiv \chi_0 F_{\text{free}} =
  \left(\frac{\chi_0}{\Tvec{v}^\dag \om{\Tvec{Z}}}\right) y.
  \label{eq:sensing_function}
\end{equation}
The factor $\Tvec{v}^\dag \om{\Tvec{Z}}/\chi_0$ is sometimes referred to as the optomechanical plant or, particularly in the context of LIGO, the sensing function~\cite{LIGOScientific:2016xax,LIGOScientific:2017aaj,Sun:2020wke}.
It is this estimated freerunning motion, or its strain-referred equivalent $h_{\text{free}} = x_{\text{free}} / L_{\text{arm}}$, which serves as the output channel for gravitational-wave detectors.}

\input{sections/signal_flow_reduced}

Next, we consider the application of feedback control. The test mass is controlled by applying a filter $C$ to the measurement $y$ to produce an external feedback force $Cy$, as shown in \cref{fig:signal_flow_reduced}. By defining the loop suppression function
\begin{equation}
  \Tmat{G}_\text{ctrl} = \left(\Tmat{1} - C\om{\Tvec{Z}}\Tvec{v}^\dag\right)^{-1},
  \label{eq:Gctrl_Hctrl}
\end{equation}
the transfer matrices describing the feedback-modified contributions to the output field $\Tvec{y}$ from the ingoing quantum vacuum field $\Tvec{s}$, from the external forces $\Fext$, from the sensing noises $\xsens$, and from any additional loss field $\Tvec{a}_\mu$ are given by, respectively,
\begin{subequations}
  \label{eq:controled_tfs}
\begin{align}
  \eff{\Tmat{H}} &= \Tmat{G}_\text{ctrl} \om{\Tmat{H}} \\
  \eff{\Tvec{Z}} &= \Tmat{G}_\text{ctrl} \om{\Tvec{Z}} \\
  \eff{\Tvec{Y}} &= \Tmat{G}_\text{ctrl} \om{\Tvec{Y}} \\
  \Tmat{T}_{\text{eff},\mu} &= \Tmat{G}_\text{ctrl} \Tmat{T}_\mu.
\end{align}
The transfer matrices describing the feedback-modified contributions to the true motion $x$ from $\Tvec{s}$, from $\Fext$, from $\xsens$, and from $\Tvec{a}_\mu$ are, respectively,
\begin{align}
  \eff{\Tvec{D}}^\dag &= \om{\Tvec{D}}^\dag + \om{\chi} C\Tvec{v}^\dag\Tmat{G}_\text{ctrl} \om{\Tmat{H}} \label{eq:controled_field_to_motion}\\
  \eff{\chi} &= \om{\chi} \left(1 + C\Tvec{v}^\dag\Tmat{G}_\text{ctrl} \om{\Tvec{Z}}\right) \\
  \eff{X} &= \om{X} + \om{\chi}C\Tvec{v}^\dag \Tmat{G}_\text{ctrl} \om{\Tvec{Y}} \label{eq:Xeff} \\
  \Tvec{J}^\dag_{\text{eff},\mu} &= \Tvec{J}^\dag_\mu + \om{\chi} C\Tvec{v}^\dag\Tmat{G}_\text{ctrl} \Tmat{T}_\mu.
\end{align}
\end{subequations}

\subsubsection{Spectral densities}
\label{subsubsec:general_spectra}

Having written down the appropriate transfer functions relating the various input fields, displacements, and forces to the true motion $x$ and measurement $y$, we are now in a position to write down the spectral densities $S_{xx}$ and $S_{yy}$.
We first tackle the appearance of quantum noise in these quantities.
As discussed in Ref.~\cite{McCuller:2021mbn}, the canonical commutation relations of the vacuum allow the spectral density of the quantum noise of the measurement to be written as\footnote{In the notation of \cite{McCuller:2021mbn}, $S_{yy}^{(\text{quant})}$ is $N\hbar\omega_0/2$.}
\begin{equation}
  \frac{S_{yy}^{(\text{quant})}}{\hbar\omega_0/2} = \left|\Tvec{v}^\dag\eff{\Tmat{H}}\Tmat{R}(\phi) \Tmat{S}(r)\right|^2
  + \sum_\mu \left|\Tvec{v}^\dag \Tmat{T}_{\text{eff},\mu}\right|^2.
\end{equation}
This expression assumes that the squeezed vacuum state $\Tvec{s}$ injected into the system is generated by taking an initially unsqueezed vacuum state and transforming it by the matrices
\begin{equation}
  \Tmat{R}(\phi) =
  \begin{bmatrix*}[r]
    \cos\phi & -\sin\phi \\
    \sin\phi & \cos\phi
  \end{bmatrix*}, \qquad
  \Tmat{S}(r) =
  \begin{bmatrix}
    \rme^{+r} & 0 \\
    0 & \rme^{-r}
  \end{bmatrix},
\end{equation}
where $r$ is the squeeze amplitude, and $\phi$ is the frequency-independent phase at which the squeezed state is injected in to the optical system.
It also assumes the quantum noise fields $\{\Tvec{a}_\mu\}$ are unsqueezed vacuum states.

Similarly, the quantum noise contribution to the true motion, from the injected squeezed state $\Tvec{s}$ as well as all sources of loss $\{\Tvec{a}_\mu\}$, is
\begin{equation}
  \frac{S_{xx}^{(\text{quant})}}{\hbar\omega_0/2} = \left|\eff{\Tvec{D}}^\dag\Tmat{R}(\phi) \Tmat{S}(r)\right|^2
  + \sum_\mu \left| \Tvec{J}_{\text{eff},\mu}\right|^2.
  \label{eq:true_motion_quantum_nofactor}
\end{equation}
As is done in Ref.~\cite{McCuller:2021mbn} for the noise of the measurement, the quantum noise contribution to the true motion can be factored into (now restoring the argument $\Omega$)
\begin{subequations}
  \label{eq:true_motion_quantum}
\begin{align}
  \frac{S_{xx}^{(\text{quant})}(\Omega)}{\hbar\omega_0/2} &= \Gamma_x(\Omega) \left[ \eta_x(\Omega) S_x(\Omega) + \Lambda_x(\Omega)\right] \\
  S_x(\Omega) &= S_- \cos^2\left[\phi + \theta_x(\Omega)\right]
  + S_+\sin^2\left[\phi + \theta_x(\Omega)\right] \\
  S_\pm &= \left[1 - \Xi'_x(\Omega)\right]\rme^{\pm 2r}
  + \Xi'_x(\Omega) \rme^{\mp 2r}. \label{eq:Spm}
\end{align}
\end{subequations}
This factorization is written in terms of motional equivalents of the McCuller metrics~\cite{McCuller:2021mbn,Ganapathy:2022hgu}:
\begin{subequations}
\label{eq:metrics}
  \begin{align}
  \theta_x(\Omega) &= \frac{1}{2} \arg \left( \frac{\xi_p + \rmi \xi_q}{\xi_p - \rmi \xi_q} \right) \label{eq:metrics_theta} \\
  \Xi_x(\Omega) &= \frac{1}{2} -
  \sqrt{
    \frac{(|\xi_p|^2 - |\xi_q|^2)^2 + 4[\real(\xi_q \xi_p^*)]^2}
    {4(|\xi_p|^2 + |\xi_q|^2)^2}
  } \label{eq:metrics_Xi} \\
  \eta_x(\Omega)\Gamma_x(\Omega) &= |\xi_p|^2 + |\xi_q|^2, \label{eq:metrics_etaGamma}
\end{align}
\end{subequations}
and
\begin{equation}
  \xi_q(\Omega) = \eff{\Tvec{D}}^\dag(\Omega)\, \Tvec{e}_q, \qquad
  \xi_p(\Omega) = \eff{\Tvec{D}}^\dag(\Omega)\, \Tvec{e}_p,
  \label{eq:xi_def}
\end{equation}
and with $\Xi'_x$ in \cref{eq:Spm} related to $\Xi_x$ in \cref{eq:metrics_Xi} in a manner described later.

The quantity $\theta_x(\Omega)$ is the intrinsic frequency-dependent phase that the squeezed state acquires by propagating through the optomechanical system, experiencing the effects of radiation pressure and cavity dispersion.
(This is distinct from $\phi$, the experimentalist-controlled phase with which the squeezed state is injected into the interferometer.)
$\Gamma_x(\Omega)$ is the optomechanical gain of the system; for a system under feedback control to cool the motion of a degree of freedom, $\Gamma_x$ will be flat in frequency below a resonance and then fall with some power of $\Omega$ above the resonance with the details depending on the optical system and the form of the feedback $C(\Omega)$.

The quantity $\Xi_x(\Omega)$ is the fundamental dephasing arising from the upper and lower sidebands experiencing different loss or the fields interacting with a lossy mechanical system.
In the case of feedback cooling, the main contribution to $\Xi_x(\Omega)$ is at frequencies around the resonance $\eff{\Omega}$ where the effective mechanical susceptibility $\eff{\chi}(\Omega)$ is most lossy. Additionally, there are technical sources of phase noises, such as the rms noise $\phi_{\text{rms}}$ of the phase between the squeezed state and the local oscillator, which are combined with this fundamental dephasing to give the total effective dephasing $\Xi'_x(\Omega)$ appearing in \cref{eq:true_motion_quantum} as described by \cref{eq:metrics_Xi'}.

The loss\,---\,i.e., the contribution to the total motion from all of the unsqueezed vacuum fields $\{\Tvec{a}_\mu\}$ entering the system\,---\,is quantified by $\Lambda_x(\Omega)$. The contribution to the motion by the squeezed vacuum injected into the interferometer is correspondingly reduced by the efficiency $\eta_x(\Omega)$.

The analogous metrics studied in Ref.~\cite{McCuller:2021mbn} for the measurement $S_{yy}^{(\text{quant})}$ are obtained by replacing $\xi_q(\Omega)$ and $\xi_p(\Omega)$ in \cref{eq:metrics} with
\begin{equation}
  m_q(\Omega) = \Tvec{v}^\dag\eff{\Tmat{H}}(\Omega)\, \Tvec{e}_q, \qquad
  m_p(\Omega) = \Tvec{v}^\dag\eff{\Tmat{H}}(\Omega)\, \Tvec{e}_p.
\end{equation}
An important consequence of this difference is that the rotation of the squeezed state as measured at the output of the interferometer $\theta_y(\Omega)$ is not the same as the rotation of the squeezed state responsible for true motion $\theta_x(\Omega)$. As a result, it is not advantageous to use the filter cavity to reduce true motion in the same way as it is used to reduce measurement noise: the filter cavity attempts to keep $\theta_y(\Omega)$, not $\theta_x(\Omega)$, equal to zero at all frequencies. Furthermore, while the motional dephasing $\Xi_x(\Omega)$ is caused by the lossy feedback, in the normal operation of gravitational wave detectors, the main contribution to the measurement dephasing $\Xi_y(\Omega)$ is at frequencies around $\sql$ where one of the sidebands is resonant in the filter cavity and experiences its losses while the other is non-resonant and does not experience the same loss. Measurement dephasing can be reduced by reducing losses; however, motional dephasing is an unavoidable consequence of the feedback damping.

The total spectral density of true motion is the sum of the quantum noise \cref{eq:true_motion_quantum} as well as the classical force and sensing noises:
\begin{equation}
  S_{xx} = S_{xx}^{(\text{quant})} + |\eff{\chi}|^2 S_{FF}^{\text{(ext)}} + |\eff{X}|^2 S_{xx}^{(\text{sens})},
  \label{eq:general_true_motion_spectrum}
\end{equation}
while the total measurement noise is
\begin{equation}
  S_{yy} = S_{yy}^{(\text{quant})} + \left|\Tvec{v}^\dag\eff{\Tvec{Z}}\right|^2 S_{FF}^{\text{(ext)}}
  + \left|\Tvec{v}^\dag \eff{\Tvec{Y}}\right|^2 S_{xx}^{(\text{sens})}.
  \label{eq:general_measurement_spectrum}
\end{equation}
These equations assume that there are no correlations between the classical noises $\Fext$ and $\xsens$, or correlations between the classical noises and the quantum noises $x_{\text{quant}}$ or $y_{\text{quant}}$.

%% file: sections/signal_flow_reduced.tex
\begin{figure}
  \centering
  \begin{tikzpicture}[
    semithick,
    ]
    \node (y) at (-2.5,0) {$y$};
    \node[dot] (y1) at (-1.25,0) {};
    \node[dot] (y2) at (0,0) {};
    \node[dot] (F1) at (4,0) {};
    \node (Fext) at (5.5,0) {$\Fext$};

    \node[dot] (x1) at (4,4) {};
    \node (x) at (5.5,4) {$x$};

    \node (xsens) at (3,1) {$\xsens$};
    \node[dot] (xsens1) at (2,2) {};

    \node (s) at (-1.5,4) {$\Tvec{s}$};
    \node[dot] (s1) at (0,4) {};

    \draw[midarrow] (y1) -- node[above] {} (y);
    \draw[midarrow] (y2) -- node[above] {$\Tvec{v}^\dagger$} (y1);

    \draw[midarrow] (F1) -- node[right] {$\chi_{\text{om}}$} (x1);
    \draw[midarrow] (Fext) -- node[above] {} (F1);
    \draw[midarrow] (x1) -- node[above] {} (x);

    \draw[midarrow] (xsens) to (xsens1);
    \draw[midarrow] (xsens1) -- node[above left] {$X_{\text{om}}$} (x1);
    \draw[midarrow] (xsens1) -- node[above left] {$\Tvec{Y}_{\text{om}}$} (y2);

    \draw[midarrow] (s) -- (s1);
    \draw[midarrow] (s1) -- node[above] {$\Tvec{D}_{\text{om}}^\dagger$} (x1);
    \draw[midarrow] (F1) -- node[below] {$\Tvec{Z}_{\text{om}}$} (y2);
    \draw[midarrow] (s1) -- node[left] {$\Tmat{H}_{\text{om}}$} (y2);

    \draw[midarrow,bend right=90,looseness=1.2] (y1) to node[above] {$C$} (F1);
  \end{tikzpicture}
  \caption{Reduced signal flow graph for the optomechanical system in \cref{fig:signal_flow_full}.
    A further reduced representation can be made by eliminating the feedback control path labeled ``$C$'' and then replacing all quantities suffixed with ``om'' by their counterparts suffixed by ``eff'' (\cref{eq:controled_tfs}).
    The additional loss fields $\{\Tvec{a}_\mu\}$ have been omitted from this graph.
  \label{fig:signal_flow_reduced}}
\end{figure}
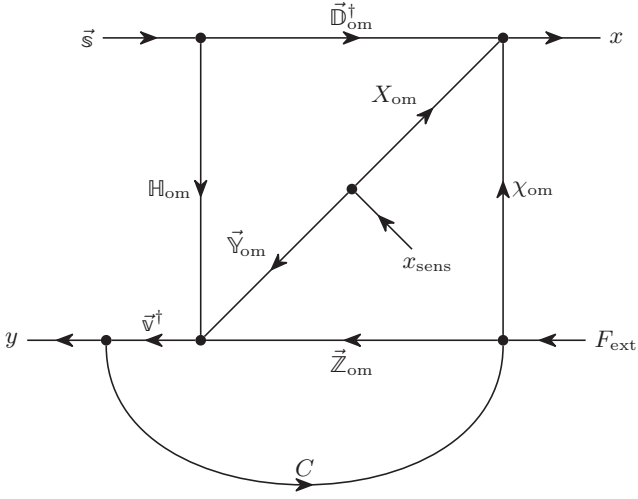

%% file: sections/single_mirror.tex
\subsection{Single free-mass mirror}
\label{subsec:single_mirror}

We now apply \cref{eq:Gctrl_Hctrl,eq:controled_tfs} in the previous section to the case of a single free-mass mirror with susceptibility $\chi_0=-1/M\Omega^2$ and amplitude reflectivity $r$ illuminated on its front surface with a laser of power $P$ and wavenumber $k$. Using a feedback control signal derived from a measurement of the optical field, we want to trap the mirror and damp its motion by engineering an effective susceptibility
\begin{equation}
  \eff{\chi} = \frac{1}{M} \times \frac{1}{\eff{\Omega}^2 - \Omega^2  + \rmi \eff{\Omega} \Omega / \eff{Q}}.
  \label{eq:vel_damped_mirror_susceptibility}
\end{equation}
This can be achieved by choosing a homodyne angle of $\zeta=0$ (the phase quadrature) to set $\Tvec{v}^\dag = \begin{bmatrix} 0 & 1 \end{bmatrix}$, and by choosing a feedback filter
\begin{equation}
  C = -\frac{M\eff{\Omega}^2}{2kr\sqrt{P}}\left(1 + \rmi \frac{\Omega}{\eff{\Omega}\eff{Q}}\right).
\end{equation}
In this scenario, we can use a number of results from \cref{sec:mirror_radiation_pressure_dynamics} for the optomechanics of a free mirror to find that when feedback is applied, the loop suppression function $\Tmat{G}_\text{ctrl}$ defined in \cref{eq:Gctrl_Hctrl} is diagonal, with its nonzero entries being 1 and $\Gctrl = \left.1\middle/\left(1 - 2 k r \sqrt{P} \chi_0 C\right)\right. = \eff{\chi} / \chi_0$.

From here we can now compute the quantities contributing to the mirror's true motion $x$.
While the effective susceptibility in \cref{eq:vel_damped_mirror_susceptibility} converts external forces to true motion,
the conversion of sensing noises to true motion is found from \cref{eq:Xeff}:
\begin{align}
  \eff{X} &= \Gctrl - 1
  = \frac{\eff{\chi}}{\chi_0} - 1 \nonumber \\
  &= - \frac{
    1 + \rmi \Omega/\eff{\Omega}\eff{Q}
  }{
    1 - (\Omega/\eff{\Omega})^2 + \rmi \Omega/\eff{\Omega}\eff{Q}
    }
    \label{eq:mirror_sens_trans}
\end{align}
This again uses results from \cref{sec:mirror_radiation_pressure_dynamics}, as well as the identification of $\om{X} = 0$, which amounts to saying that in the absence of an optical cavity or measurement-based feedback, there is no way for a sensing noise $\xsens$ to affect the center of mass motion $x$ of the mirror.

In order to compute the quantum noise and the McCuller metrics \cref{eq:metrics}, we now compute $\xi_q$ and $\xi_p$, the two components of the feedback-modified field-to-displacement transfer function $\Tvec{D}_{\text{eff}}^\dagger$. Using the results of \cref{sec:mirror_radiation_pressure_dynamics} with \cref{eq:controled_field_to_motion}, we find
\begin{subequations}
\label{eq:mirror_motion_trans}
\begin{align}
  \xi_q &= \chi_0 \left[ \frac{2(2\mathcal{R} + \mathcal{L}) \sqrt{P}}{c} + r\Krp C\Gctrl\right] \label{eq:mirror_xi_q}\\
  \xi_p &= -r \chi_0 C\Gctrl, \label{eq:mirror_xi_p}
\end{align}
\end{subequations}
with $\mathcal{R} = r^2$ being the mirror's power reflectivity, $\mathcal{L}$ being the power lost at the mirror (i.e., neither reflected nor transmitted), and $c$ being the vacuum speed of light.
$\mathcal{K}$ is an optomechanical coupling defined below.
In the absence of feedback ($C=0$), true mirror motion is driven only by vacuum fluctuations in the amplitude quadrature, since $\xi_p=0$. However, when the measurement $y$ is formed by reading out the phase quadrature and is fed back to the mirror, vacuum fluctuations in that quadrature are impressed on the test mass (\cref{eq:mirror_xi_p}). Furthermore, since radiation pressure converts amplitude fluctuations into phase fluctuations, amplitude quadrature fluctuations impart true motion through the feedback (the second term in \cref{eq:mirror_xi_q}) in addition to their direct coupling (the first term in \cref{eq:mirror_xi_q}). This optomechanical coupling between amplitude and phase quadratures is given by \cref{eq:general_rp_coupling}, which for a free mirror is
\begin{equation}
  \Krp = - \frac{4k(2\mathcal{R} + \mathcal{L})P}{cM\Omega^2} = - \left(\frac{\sql}{\Omega}\right)^2,
\end{equation}
where the SQL frequency $\sql$ is the frequency where radiation pressure and shot noise produce equal contributions to the measurement $y$; here, it is given by
\begin{equation}
  \label{eq:W_sql}
  \sql^2 = \frac{4k(2\mathcal{R} + \mathcal{L})P}{cM}.
\end{equation}
Now substituting our choice of $C$ into \cref{eq:mirror_motion_trans}, the two components of $\Tvec{D}_\text{eff}^\dag$ (\cref{eq:xi_def}) are
\begin{subequations}
  \label{eq:xi_velocity_damped}
  \begin{align}
    \xi_q &= \frac{1}{2k\sqrt{P}} \times
    \frac{
      (\sql/\eff{\Omega})^2
    }{
      1 - (\Omega/\eff{\Omega})^2 + \rmi \Omega/\eff{\Omega}\eff{Q}
    } \\
    \xi_p &= \frac{1}{2k\sqrt{P}} \times
    \frac{
      1 + \rmi \Omega/\eff{\Omega}\eff{Q}
    }
    {
      1 - (\Omega/\eff{\Omega})^2 + \rmi \Omega/\eff{\Omega}\eff{Q}
    }.
  \end{align}
\end{subequations}

The frequency at which the contribution to the true motion from radiation pressure is equal to the contribution from shot noise\,---\,i.e., the frequency for which $|\xi_q| = |\xi_p|$\,---\,we denote by $\Omega_x$. It only exists if $\eff{\Omega} < \sql$ and is given by
\begin{equation}
  \Omega_x^2 = \eff{Q}^2\eff{\Omega}^2
  \left[\left(\frac{\sql}{\eff{\Omega}}\right)^4 - 1 \right].
  \label{eq:W_x}
\end{equation}
The rotation of the squeezed state responsible for true motion (\cref{eq:metrics_theta}) is
\begin{equation}
  \theta_x(\Omega) = \frac{1}{2}
  \arctan \left[\frac{
    2(\sql/\eff{\Omega})^2
  }{
    (\Omega/\eff{\Omega}\eff{Q})^2 - (\Omega_x/\eff{\Omega}\eff{Q})^2
  }\right].
\end{equation}
\begin{widetext}
\noindent%
The optomechanical gain (\cref{eq:metrics_etaGamma}) is
\begin{equation}
  \eta_x(\Omega)\Gamma_x(\Omega) = \frac{1}{4 k^2 P} \times \frac{
    2 + (\Omega/\eff{\Omega}\eff{Q})^2 + (\Omega_x/\eff{\Omega}\eff{Q})^2
  }{
    \left[1 - (\Omega/\eff{\Omega})^2\right]^2 + (\Omega/\eff{\Omega}\eff{Q})^2
  }
\end{equation}
and the dephasing (\cref{eq:metrics_Xi}) is
\begin{equation}
  \Xi_x(\Omega) = \frac{1}{2}\left\{1 -
  \frac{
    \sqrt{\left[(\Omega/\eff{\Omega}\eff{Q})^2 - (\Omega_x/\eff{\Omega}\eff{Q})^2\right]^2 + 4\left[1 + (\Omega_x/\eff{\Omega}\eff{Q})^2\right]}
  }{
    2 + (\Omega/\eff{\Omega}\eff{Q})^2 + (\Omega_x/\eff{\Omega}\eff{Q})^2
  }\right\}.
\end{equation}
Note that the dephasing vanishes in the limit of no damping ($\eff{Q}\to\infty$).

The foregoing equations, with some additional specification of loss $\Lambda_x$ and rms phase noise, are enough to define the spectrum $S_{xx}^{\text{(quant)}}(\Omega)$ of the true quantum-noise-induced motion via \cref{eq:true_motion_quantum,eq:metrics}.
Then the total spectrum of the true motion (\cref{eq:general_true_motion_spectrum}), incorporating classical force and sensing noises, is
\begin{equation}
  S_{xx}(\Omega) = S_{xx}^{(\text{quant})}(\Omega)
  + \frac{1}{\left[1 - (\Omega/\eff{\Omega})^2\right]^2 + (\Omega/\eff{\Omega}\eff{Q})^2}
  \Bigg\{
    \frac{1}{M^2\eff{\Omega}^4} S_{FF}^{\text{(ext)}}(\Omega)
    + \left[1 + \left(\frac{\Omega}{\eff{\Omega}\eff{Q}}\right)^2\right]S_{xx}^{(\text{sens})}(\Omega)
    \Bigg\}.
  \label{eq:Sxx}
\end{equation}
In the particular case of no loss ($\Lambda_x = \mathcal{L} = 1-\mathcal{R} = 0$), no rms phase noise, and frequency-independent squeezing with squeeze factor $r$ and squeeze angle $\phi$, the quantum noise is
\begin{equation}
  \label{eq:Sxx_quantum_lossless_reduced}
  S_{xx}^{\text{(quant)}}(\Omega) = \frac{\hbar}{M \sql^2} \times \frac{\rme^{+2r} (\sql / \eff{\Omega})^4 + \rme^{-2r} [1 + (\Omega/\eff{\Omega}\eff{Q})^2]}{[1 - (\Omega/\eff{\Omega})^2]^2 + (\Omega/\eff{\Omega}\eff{Q})^2}.
\end{equation}
\end{widetext}
We note that \cref{eq:Sxx_quantum_lossless_reduced} corresponds precisely to the expression
\begin{equation}
  |\eff{\chi}|^2 S_{FF}^{\text{(rp)}} + |\eff{X}|^2 S_{xx}^{\text{(shot)}}
  \label{eq:semi-classical_quantum}
\end{equation}
with
\begin{align}
  S_{FF}^{\text{(rp)}} &= 8\hbar \omega_0 P \rme^{+2r} / c^2 = \hbar \rme^{+2r} M \sql^2 \\
  S_{xx}^{\text{(shot)}} &= \hbar c^2 \rme^{-2r} / 8 \omega_0 P = \hbar \rme^{-2r} / M \sql^2;
\end{align}
and $\eff{\chi}$ given by \cref{eq:vel_damped_mirror_susceptibility} and $\eff{X}$ given by \cref{eq:mirror_sens_trans}.
In other words, the above derivation for the quantum noise of a single lossless mirror coincides exactly with what one would expect by treating the quantum noise semi-classically as a sum of radiation-pressure and shot-noise fields entering the system, as anticipated in \cref{sec:big_picture}. Comparing with \cref{eq:general_true_motion_spectrum}, we also see that the quantum radiation pressure and shot noise effectively enter as classical force and sensing noises, respectively, in this simple case.

%% file: sections/rse_ifo.tex
\subsection{Interferometer}
\label{subsec:simplified_ifo}

We now consider the dual recycled Fabry--Perot Michelson interferometer, specializing to the case where the signal extraction cavity is tuned to broaden the intrinsic bandwidth of the arm cavities. This technique, known as resonant sideband extraction (RSE)~\cite{Mizuno:1993cj}, is used by LIGO and is planned to be used for Cosmic Explorer. As shown in \cref{fig:signal_flow_full}, this is modeled by treating the differential motion of the two arm cavities as a single effective arm cavity formed by one end test mass (ETM) and one input test mass (ITM). The differential mode of the full interferometer is then equivalent, up to constant factors to be explained, to a three-mirror coupled cavity where the addition of the signal extraction mirror (SEM) forms the signal extraction cavity.

In an instrument with this topology, the relevant degree of freedom is the differential motion of the four test masses, viz.
\begin{equation}
  x_- = \left(x_{\text{EX}} - x_{\text{IX}}\right) - \left(x_{\text{EY}} - x_{\text{IY}}\right),
  \label{eq:darm}
\end{equation}
which is equivalent to twice the motion of the mirrors in a single effective arm cavity. Both mirrors in each arm cavity are free to move with a bare susceptibility $\chi_0(\Omega)$; however, in the case of gravitational wave detectors the mirrors are made highly reflective and we will be interested in dynamics slower than the light travel time between the mirrors, so, to a good approximation, this system is equivalent to one in which the ITM is fixed and the dynamics of the cavity are described entirely by the ETM with susceptibility $2\chi_0(\Omega)$. In the case of a free mass, which we continue to study here, this is thus equivalent to an arm cavity having one free mirror with effective mass $M / 2$, as depicted in \cref{fig:signal_flow_full}.

We first describe the relevant dynamics of the interferometer before feedback is applied~\cite{Buonanno:2001cj,McCuller:2021mbn}. To do so, one can start with the radiation pressure dynamics of the ETM described in \cref{sec:mirror_radiation_pressure_dynamics}, using $\chi_0=-2/M\Omega^2$ and the arm power $P_\text{a}$ for $P$ in the optomechanical coupling $\Krp$ of \cref{eq:general_rp_coupling}, to reduce the ETM portion of the interferometer graph in \cref{fig:signal_flow_full}. It is useful to describe the optical propagation throughout the rest of the system in terms of the transfer functions for the upper and lower sidebands in the absence of radiation pressure. The reflection of optical fields from the dark port of the interferometer is described by the transfer function $\rrse(\Omega)$ and the transmission of signals from the end test mass through the interferometer to the dark port is $\trse(\Omega)$; approximate algebraic expressions for these quantities will be given later. The reflection of the two-photon optical fields from the interferometer can then be written as
\begin{equation}
  \mrse{\Tmat{H}} = -\rrse\Tmat{1} + r_\text{e} \trse^2\Krp \Tvec{e}_p\Tvec{e}_q^\dag.
  \label{eq:rse_reflection}
\end{equation}
The first term indicates that the amplitude and phase quadratures incident on the dark port reflect off the interferometer with a factor $\rrse$. The second term indicates that the amplitude fluctuations also propagate to the ETM, are converted to phase fluctuations through the optomechanical coupling, and propagate back to the dark port. This ponderomotive coupling of amplitude to phase is thus the same as that of a single mirror, but modified by two factors of $\trse$.

Amplitude quadrature fluctuations entering the dark port drive true test mass motion after propagating once through the interferometer. The two arms of the full interferometer result in twice the motion of a single arm; however, the presence of the beamsplitter in the interferometer reduces the amplitude of one-way propagation by a factor of $\sqrt{2}$, resulting in
\begin{equation}
  \mrse{\Tvec{D}}^\dag = \frac{\trse}{\sqrt{2}}\times 2 \Tvec{D}^\dag_\text{mirror}
  = \sqrt{2}\times\frac{2(2\mathcal{R} + \mathcal{L})\chi_0\sqrt{P_\text{a}}}{c}\trse \Tvec{e}_q^\dag,
  \label{eq:rse_field_to_motion}
\end{equation}
where $\Tvec{D}^\dag_\text{mirror}$ is given by \cref{eq:mirror_force_to_motion}; here again $\mathcal{R} = r_\text{e}^2$ is the power reflectivity of the end test mass, and $\mathcal{L}$ is its power loss. External force $\Fext$ also produces phase fluctuations at the dark port after propagating once through the interferometer, again reduced by a factor of $\sqrt{2}$ because of the beamsplitter:
\begin{equation}
  \mrse{\Tvec{Z}} = \frac{\trse}{\sqrt{2}}\Tvec{Z}_\text{mirror}
  = \frac{1}{\sqrt{2}} \times 2kr\chi_0\sqrt{P_\text{a}}\trse \Tvec{e}_p,
  \label{eq:rse_motion_to_field}
\end{equation}
where $\Tvec{Z}_\text{mirror}$ is given by \cref{eq:mirror_motion_to_field}. Since no cavities are detuned from resonance, which would mix amplitude and phase quadratures, the optomechanical susceptibility is unmodified ($\mrse{\chi} = \chi_0$) and there is no conversion of sensing noises into true motion ($\mrse{X} = 0$).

After feedback is applied, the transfer function from fields to true motion (\cref{eq:xi_def}) are thus, by \cref{eq:controled_field_to_motion},
\begin{subequations}
  \label{eq:rse_field_to_motion_with_feedback}
\begin{align}
  \xi_q &= \sqrt{2}\chi_0 \left[ 2(2\mathcal{R} + \mathcal{L}) \frac{\sqrt{P_\text{a}}}{c}\trse + r_{\text{e}} C\Gctrl \Krp\trse^2 \right] \\
  \xi_p &= -\sqrt{2}\chi_0 C \Gctrl \rrse.
\end{align}
\end{subequations}

Further insight can be gained by looking at the form of the sideband transfer functions, which can be approximated as
\begin{align}
  \rrse(\Omega) &= \frac{1 - \rmi \Omega/\Omega_\text{rse}}{1 + \rmi \Omega/\Omega_\text{rse}}
    \underset{\Omega\ll \Omega_\text{rse}}{\approx} 1 \label{eq:rrse_approx} \\
  \trse(\Omega) &= \sqrt{\frac{\Fa}{\Fs}}
  \times\frac{1}{1 + \rmi \Omega/\mrse{\Omega}}
  \underset{\Omega\ll \mrse{\Omega}}{\approx} \sqrt{\frac{\Fa}{\Fs}}, \label{eq:trse_approx}
\end{align}
where $\mrse{\Omega}$ is the RSE pole, given by
\begin{equation}
  \mrse{\Omega} = \frac{1 + r_{\text{s}}}{1 - r_{\text{s}}} \times \Omega_{\text{a}}, \qquad
  \Omega_{\text{a}} = \frac{c}{L_{\text{a}}}\times\frac{1 - r_{\text{i}}}{1 + r_{\text{i}}}.
\end{equation}
Here $L_{\text{a}}$ is the length of the arm, $r_{\text{i}}$ is the amplitude reflectivity of the input test mass, $r_\text{s}$ is the amplitude reflectivity of the signal extraction mirror, and $\Fa$ and $\Fs$ are the finesses of the arm and signal extraction cavities, respectively:
\begin{equation}
  \Fa = \frac{\pi}{1 - r_{\text{i}}}, \qquad
  \Fs = \frac{\pi}{1 - r_{\text{s}}};
\end{equation}
here we have additionally assumed $|r_\text{e}| \ll |r_\text{i}|$, which is almost always the choice adopted in gravitational-wave interferometers with a Fabry--Perot Michelson topology.
The approximate sideband transfer functions \cref{eq:rrse_approx,eq:trse_approx} are only qualitatively accurate for Cosmic Explorer; however, in the case of feedback cooling we will always be interested in frequencies $\Omega \ll \mrse{\Omega}$ where they are valid for all cases considered here.

From \cref{eq:rse_reflection,eq:rse_field_to_motion,eq:rse_motion_to_field,eq:rse_field_to_motion_with_feedback} we see that each factor of $\sqrt{P_\text{a}}$ is accompanied by a factor of $\trse$. Thus, comparing \cref{eq:rse_field_to_motion_with_feedback} with \cref{eq:mirror_motion_trans}, we see that, in the $\Omega \ll \mrse{\Omega}$ limit, the dynamics are the same as those of a mirror after replacing $P$ with $P_\text{a}\Fa/\Fs$.
Using an SQL frequency of
\begin{equation}
  \left[\sql^{(\text{rse})}\right]^2 = \frac{\Fa}{\Fs}\frac{4k(2\mathcal{R} + \mathcal{L})}{cM/2} P_\text{a},
  \label{eq:rse-W_sql}
\end{equation}
the components of $\mrse{\Tvec{D}}^\dag$ are thus
\begin{equation}
  \xi_{p,q}^{(\text{rse})} = \sqrt{\frac{2\Fs}{\Fa}}\xi_{p,q}^{(\text{mirror})},
\end{equation}
where $\xi_{p,q}^{(\text{mirror})}$ is given by \cref{eq:xi_velocity_damped}.
Therefore, all of the results of \cref{subsec:single_mirror} are the same except that the optomechanical gain is scaled by a factor of $2\Fs/\Fa$:
\begin{subequations}
  \label{eq:rse-McCuller-metrics}
  \begin{align}
    \theta_x^{(\text{rse})}(\Omega) &\simeq \theta_x^{(\text{mirror})}(\Omega) \\
    \Xi_x^{(\text{rse})}(\Omega) &\simeq \Xi_x^{(\text{mirror})}(\Omega) \\
    [\eta_x(\Omega)\Gamma_x(\Omega)]^{(\text{rse})} &\simeq \frac{2\Fs}{\Fa}[\eta_x(\Omega)\Gamma_x(\Omega)]^{(\text{mirror})}.
  \end{align}
\end{subequations}
We stress that the McCuller motion metrics given by \cref{eq:rse-McCuller-metrics} have the same form as the corresponding metrics of \cref{subsec:single_mirror} but are to be evaluated with the RSE SQL frequency given by \cref{eq:rse-W_sql} rather than that given by \cref{eq:W_sql}.

%% file: sections/thermometry.tex
\section{Thermometry and occupation number}
\label{sec:thermometry}

In \cref{sec:true_motion} we worked out the appearance of quantum and classical noises in the true motion of a single mirror, and subsequently of the differential motion of a four-test-mass interferometer.
We now discuss the relation of such motion to the motion of a harmonic oscillator at finite temperature, which is the premise of optomechanical feedback-cooling experiments.
We assume that we have a spectral estimate $S_{xx}(\Omega)$ of the physical motion $x$ of the mirror motion, or the differential interferometer motion.
The question now is how to turn this motion into an effective temperature $\osc{T}$ or occupation number $\osc{n}$.
One procedure is to compare the motional spectrum to the theoretical spectrum of a thermally-limited oscillator, calculated from the fluctuation--dissipation theorem~\cite{Aspelmeyer:2013lha}:
\begin{align}
    \label{eq:ideal_thermal}
    S_{xx}^{\text{(osc)}}(\Omega)
    &= 2\hbar \coth\left(\frac{1}{2}\frac{\hbar \Omega}{k_{\text{B}} \osc{T}}\right) \, \bigl\lvert\imag{\osc{\chi}{(\Omega)}}\bigr\rvert \\
    \label{eq:fdt_quantum}
    &= 4\hbar \left(\osc{n}(\Omega) + \tfrac{1}{2}\right)  \bigl\lvert\imag{\osc{\chi}{(\Omega)}}\bigr\rvert,
\end{align}
with the frequency-dependent occupation factor $\osc{n}(\Omega) = 1/(\rme^{\hbar\Omega/k_{\text{B}} \osc{T}} - 1)$.
In the previous measurement-based feedback trapping and damping experiments on LIGO, the relevant susceptibility $\osc{\chi}$ has been taken to be a damped oscillator with resonance frequency equal to the trap frequency $\eff{\Omega}$, a damping rate $\eff{\Omega}/\eff{Q}$, and a mass $\mu = M/4$, with $M$ being the mass of a single test mass.
For such an oscillator, the power spectrum of its thermal noise at temperature $\osc{T}$ from \cref{eq:ideal_thermal} is equal to the incoherent sum of the thermal noises of four oscillators also at temperature $\osc{T}$, each with susceptibility $\osc{\chi}/4$\,---\,i.e., each with a mass $M$.
We will proceed with this convention, although we note that, regardless of the choice of $\mu$, the total noise of the true differential motion $S_{xx}(\Omega)$ does not generally have the functional form of a thermally-limited oscillator, due to the presence of quantum noise $S_{xx}^{\text{(quant)}}$, as well as classical force or sensing noises.
Mathematically, this can be handled by letting $\osc{T}$ be frequency-dependent~\cite{Clerk:2008tlb}, but it points to the necessity of either restricting the analysis to a single frequency or performing an averaging procedure in order to arrive at a single number characterizing the oscillator temperature.

A simple form of thermometry is to examine the true spectrum $S_{xx}(\Omega)$ at the resonance $\Omega_{\text{eff}}$ and compare it to $S_{xx}^{\text{(osc)}}{(\Omega_{\text{eff}})}$ as given by \cref{eq:fdt_quantum}.
To get intuition, we can examine specifically the case of a single velocity-damped mirror, whose true motion is given by \cref{eq:Sxx}.
We also recall the velocity-damped susceptibility, \cref{eq:vel_damped_mirror_susceptibility}, and note that in the Laplace domain, the frequency dependence of this susceptibility comprises two poles,
\begin{equation}
    s_{\pm} = \frac{\eff{\Omega}}{2\eff{Q}} \left(-1 \pm \sqrt{1 - 4\eff{Q}^2}\right).
    \label{eq:susceptibility_poles}
\end{equation}
When the system is critically damped ($\eff{Q} = 1/2$) or underdamped ($\eff{Q} > 1/2$), these poles both have magnitude $|s_\pm| = \eff{\Omega}$; thus, by examining the spectrum at the Fourier frequency $\eff{\Omega}$, one is examining the system's noise properties exactly on resonance.
For simplicity we specialize to the case where the light has a squeeze factor $r$, and a squeeze angle $\phi = 0$.
If the system is lossless and has no excess quantum-mechanical phase noise, the relevant quantum noise contribution is given by \cref{eq:Sxx_quantum_lossless_reduced}, and the occupation number $\osc{n}$ at $\eff{\Omega}$ is then found to be
\begin{multline}
  \label{eq:point_thermometry}
  4\left(\osc{n}{(\eff{\Omega})}+\tfrac{1}{2}\right) = \eff{Q} \left(\frac{\rme^r\, \sql}{\eff{\Omega}}\right)^2 \\
  + \left(\eff{Q} + \frac{1}{\eff{Q}}\right) \left(\frac{\eff{\Omega}}{\rme^r\, \sql}\right)^2 \\
  + \frac{S_{FF}^{\text{(ext)}}{(\eff{\Omega})}}{M^2 \eff{\Omega}^4} + \left(1 + \frac{1}{\eff{Q}^2}\right) S_{xx}^{\text{(sens)}}{(\eff{\Omega})}.
\end{multline}
The first two terms arise from the quantum noise, while the final two terms are external force noise and sensing noise.
By extremizing this formula in the presence of quantum noise alone, one finds that for a fixed value of $\eff{\Omega} / \rme^{r} \sql{}$, the optimal $\eff{Q}$ according to this single-frequency thermometry is $\left.1\middle/\sqrt{(\rme^{r}\,\sql{} / \eff{\Omega})^4 + 1}\right.$.
Substituting this optimal $\eff{Q}$ into \cref{eq:point_thermometry}, one concludes that ground state cooling is possible, in the sense of attaining $\osc{n}(\eff{\Omega}) < 1$, when $\eff{\Omega}/\rme^r\,\sql{} < 2^{3/4} \approx 1.7$.
In particular, adjusting the feedback control so as to place $\eff{\Omega} = \rme^{r}\,\sql{}$, the optimal $\eff{Q}$ is $\left.1\middle/\sqrt{2}\right.$, and the resulting occupation is $\osc{n}(\eff{\Omega}) = \left.\left(\sqrt{2} - 1\right)\middle/2\right. \approx 0.2$\,---\,hence, one would conclude from this thermometry scheme that ground state cooling is possible so long as the condition $\eff{\Omega} = \rme^{r} \sql{}$ is experimentally realizable.
Generally, in a gravitational-wave interferometer one places $\sql{}$ (via choice of power $P$ or mass $M$ or cavity finesse) in or near the desired band of gravitational-wave observation, so that the measurement is not excessively dominated by either amplitude or phase fluctuations of the quantum field at the dark port.
At the same time, the bandwidth of force actuation on the test masses typically extends from dc and into the lower end of the observation band, to provide the ability to suppress force fluctuations.
These basic features explain why in Advanced LIGO it has been possible to create an effective oscillator frequency in the vicinity of the squeezing-modified SQL frequency, and why it is likely to be possible in Cosmic Explorer as well.
Obviously, the presence of the latter two terms in \cref{eq:point_thermometry}\,---\,i.e., classical force and sensing noise\,---\,can complicate the situation by degrading the amount of cooling, which may motivate a choice of a different $\eff{\Omega}$, or of a particular squeeze amplitude $r$.
We also note that $\phi = 0$ is not the optimal choice for achieving a minimum occupation; we will return to this point in the context of the full interferometers in \cref{sec:parameter_choice}.

One might also ask if it is advantageous to choose $\eff{\Omega}$ to be significantly smaller than $\rme^r \sql{}$.
Indeed, by allowing $\eff{\Omega}$ to vary relative to $\rme^{r}\,\sql{}$ while adjusting $\eff{Q}$ to its optimal value, one finds $\osc{n}(\eff{\Omega}) \rightarrow 0$ as $\eff{\Omega} / \rme^{r} \sql{} \rightarrow 0$; i.e., in the quantum-limited case, single-frequency thermometry can indicate an arbitrarily small occupation number by choosing $\eff{\Omega} / \rme^{r} \sql{}$ to be sufficiently small.
However, this requires an overdamped system ($\eff{Q} < 1/2$), where the system's two poles (\cref{eq:susceptibility_poles}) no longer both have magnitude $\eff{\Omega}$; rather, in the strongly overdamped limit one has $s_+ \rightarrow -\eff{\Omega} \eff{Q}$ and $s_- \rightarrow -\eff{\Omega} / \eff{Q}$; thus as $Q \rightarrow 0$, these two poles approach $0$ and $-\infty$, respectively.
In the overdamped regime, therefore, the Fourier frequency $\eff{\Omega}$ no longer corresponds to the dynamical frequencies of the optomechanical system, and thus one can no longer interpret $\osc{n}(\eff{\Omega})$ as the system's on-resonance occupation number; instead, the noise power of the oscillator's motion is spread over a wide bandwidth in the Fourier domain.

Rather than a single-frequency estimate, one may also construct an occupation number estimate by integrating the motional spectrum $S_{xx}(\Omega)$.
If the interval of integration is sufficiently wide so as to capture most of the noise power of the oscillator motion, this integrated power can be compared directly to the mean-square zero point motion $\zpf^2 = \hbar/2\mu\eff{\Omega}$, which is the procedure undertaken in, e.g., Ref.~\cite{Rossi:2018dav}.
For experiments in the audio band, such as gravitational-wave detectors, a variety of classical noises can dominate the total motional spectrum at frequencies comparable to the oscillator's frequency $\eff{\Omega}$.
In particular, classical force noises such as seismic noise generically dominate the kilometer-scale gravitational-wave interferometers at frequencies $\Omega \lesssim 2\pi\times\qty{10}{\Hz}$.
For cooling experiments with such apparatus, this has motivated bandlimited computations of the occupation numbers (see, e.g., Refs.~\cite{LIGOScientific:2009mif,Whittle:2021mtt} for Initial and Advanced LIGO, and Ref.~\cite{Corbitt:2007spn} for a similar experiment with a tabletop apparatus).

Similar to previous computations with gravitational-wave detector cooling experiments, we perform a bandlimited computation of the mean-squared power over a finite interval $[\Omega_-,\Omega_+]$.
We compare this to the mean-squared power of the zero-point motion over the same interval in order to arrive at an effective occupation number $\osc{\overline{n}}$:
\begin{equation}
  \label{eq:neff_quantum}
  4\hbar\left(\osc{\overline{n}} + \tfrac{1}{2}\right) = \frac{\int_{\Omega_-}^{\Omega_+} (\rmd\Omega / 2\pi) S_{xx}(\Omega)}
  {\int_{\Omega_-}^{\Omega_+} (\rmd\Omega/2\pi) \bigl\lvert\imag{\osc{\chi}(\Omega)}\bigr\rvert}.
\end{equation}
Effectively, this yields a frequency-independent average of $\osc{n}(\Omega)$ on the interval of integration, which can be related to a single oscillator temperature $\osc{\overline{T}}$ by $\osc{\overline{n}} = 1/(\rme^{\hbar\eff{\Omega}/k_{\text{B}}\osc{\overline{T}}}-1)$.\footnote{Alternatively, we could have solved for the temperature $\osc{T}$ such that the integral of \cref{eq:ideal_thermal} over $[\Omega_-,\Omega_+]$ matches the total mean-squared power of the motional spectrum $S_{xx}(\Omega)$ integrated over the same interval, and hence compute an effective occupation number via the on-resonance Bose occupation $1/[\exp{(\hbar\eff{\Omega}/k_\text{B}\osc{T})} - 1]$.
  In \cref{sec:ligo_ce} we nonetheless find that the $\osc{\overline{n}}$ and $\osc{\overline{T}}$ found via \cref{eq:neff_quantum} produces a thermal spectrum that closely matches the total budgeted motion spectrum over the interval of interest.
  Regardless of the particular method of arriving at a temperature or occupation number, for on-resonance occupation numbers far below unity, the dependence of \cref{eq:ideal_thermal} on temperature (or occupation number) is only evident at frequencies far below $\eff{\Omega}$, and therefore outside the prescribed integration interval.
  Evidently, thermometry via bandlimited integration of the motional power spectral density is not a robust procedure for determining occupation numbers far below unity.
\label{fn:thermometry}
}
Finally, we must choose the domain $[\Omega_-,\Omega_+]$.
A reasonable requirement is that the integral of \cref{eq:fdt_quantum} over the interval should contain a majority of the thermal noise power.
This can be satisfied by choosing $\Omega_\pm = \eff{\Omega} (1 \pm 1/2\eff{Q})$.
This choice of interval will contain at least half the noise power for an underdamped or critically damped oscillator ($\eff{Q} \ge 1/2$).
Capturing at least half the noise power of an overdamped oscillator ($\eff{Q} < 1/2$) would require an even wider interval of integration.
Eventually, the lower limit of integration encounters the classical force noises that dominate over the quantum noises, and one will find an occupation number greatly in excess of the quantum limit.
As we shall see in the next section, it is possible to achieve a bandlimited $\osc{\overline{n}} < 1$ according to \cref{eq:neff_quantum} with mildly underdamped oscillators.

%% file: sections/ligo_ce.tex
\section{Application to LIGO and Cosmic Explorer}
\label{sec:ligo_ce}

\begin{figure*}[t]
  \centering
  \includegraphics[width=\textwidth]{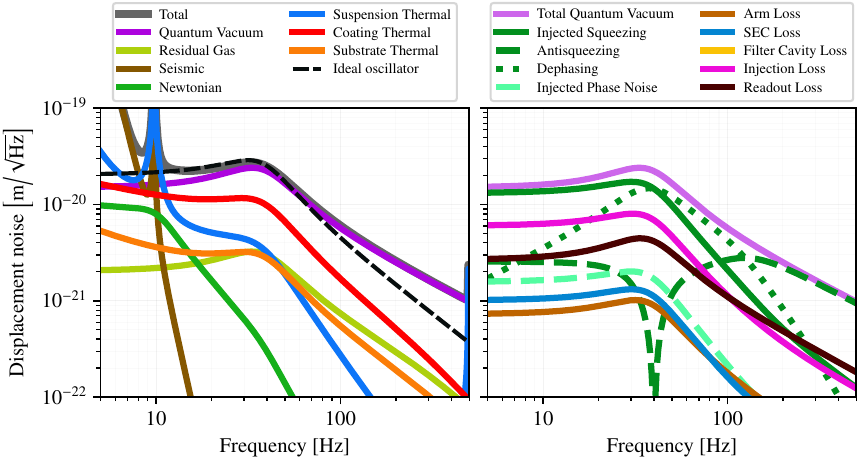}
  \caption{
    An example budget of the physical test mass motion for an oscillator in LIGO A+ trapped and cooled with measurement-based feedback control, with frequency-independent squeezed light (no filter cavity).
    The frequency-independent squeeze angle $\phi$ is adjusted to counteract the squeezed-state rotation $\theta_x$ near the resonance frequency $\eff{\Omega}$, which results in the null in the budgeted antisqueezing.
    The injected squeeze amplitude $r$ is tuned so that the dephasing, which increases with $r$, is roughly equal to the injected squeezing, which decreases with $r$.
    The two largest contributions to the true motion $x$ near the resonance are the quantum noise entering the detector's dark port and the classical sensing noise from the Brownian motion of the test mass coatings.
    Referring back to \cref{fig:signal_flow_reduced}, the quantum noise, labeled by $\Tvec{s}$, contributes to the true motion via the transfer function $\eff{\Tvec{D}^\dag}$, while the coating noise, part of the total sensing noise $x_\text{sens}$, contributes to the true motion via the transfer function $\eff{X}$.
    These are distinct from the transfer functions ($\Tvec{v}^\dag \eff{\Tmat{H}}$ and $\Tvec{v}^\dag \eff{\Tvec{Y}}$, respectively) that propagate these noises into the measurement record $y$, which is in turn used to characterize the interferometer sensitivity when used for detecting gravitational waves; therefore, the relative magnitudes of the noises that form the budget of the true motion $x$ differ in general from the relative magnitudes of the noises in a budget of the measurement $y$.
  \label{fig:aplus_physical_fis}
  }
\end{figure*}

The formalism developed in \cref{sec:true_motion} gives us the ability to model quantitatively the quantum and classical noises appearing in LIGO and Cosmic Explorer, and to assess the achievable performance of hypothetical feedback cooling experiments.
In brief, both of these instruments are dual recycled Fabry--Perot Michelson interferometers.
LIGO's test masses are each \qty{40}{\kg} cylinders of room-temperature fused silica, and each arm is \qty{4}{\km} long, with up to \qty{800}{\kW} of \qty{1064}{\nm} laser light power~\cite{LIGOScientific:2014pky}.
A possible future version of the LIGO detector, called LIGO Voyager, could use heavier cryogenic silicon test masses with \qty{2}{\um} laser light and more optical power~\cite{LIGO:2020xsf}.
The Cosmic Explorer concept~\cite{Evans:2021gyd} calls for \qty{40}{\km} of arm length with \qty{1.5}{\MW} of \qty{1064}{\nm} laser light power and \qty{320}{\kg} room-temperature fused silica mirrors.
An upgrade to Cosmic Explorer using Voyager-style cryogenic technology could increase the power to \qty{12}{\MW}.
\cref{sec:budgeted_noises} describes how the various noises, particularly the quantum noise, are folded into budgets of the differential test mass motion of these detectors.
\cref{sec:parameter_choice} then discusses how the operating parameters for the detectors are chosen in the context of a feedback cooling experiment, and \cref{sec:results} presents the resulting occupation numbers for hypothetical feedback cooling experiments using the thermometry conventions described in \cref{sec:thermometry}.

\subsection{Budgeted noises}
\label{sec:budgeted_noises}

As we have remarked in previous sections, noise budgets of gravitational-wave interferometers are typically given in terms of equivalent freerunning displacement, which is a filtered version of the measurement $y$ so defined in \cref{eq:general_measurement_spectrum}.
Instead, in this section we are primarily concerned with budgets of the true motion $x$ (\cref{eq:darm,eq:general_true_motion_spectrum}).

The transfer functions defined in \cref{eq:controled_tfs} are calculated for the dual recycled Fabry--Perot Michelson interferometer shown in \cref{fig:signal_flow_full}, without the approximations of \cref{subsec:simplified_ifo}, using the \texttt{wield.control} package~\cite{ref:wield}. The spectrum of true motion is then calculated using \cref{eq:general_true_motion_spectrum}. In particular, the quantum noise $S_{xx}^{\text{(quant)}}$ is given by \cref{eq:true_motion_quantum}.
We decompose $S_{xx}^{\text{(quant)}}$ into several terms.
The first is the \emph{injected squeezing}, which refers to main contribution of the squeezed quadrature of the field $\Tvec{s}$ injected into the interferometer\,---\,either sent towards the filter cavity if frequency-dependent squeezing is employed, or sent directly towards the interferometer if it is not.
Its power spectral density is given by
  \begin{equation}
    \eta_x(\Omega) \left[1 - \Xi'_x(\Omega)\right] \rme^{-2r} \cos^2{\left[\phi + \theta_x(\Omega)\right]},
    \label{eq:injected_squeezing}
  \end{equation}
where $\phi$ is the ingoing squeeze angle into the interferometer's dark port, $r$ is the ingoing squeeze amplitude, $\theta_x(\Omega)$ is the rotation of the squeezed state when it arrives at the test masses (\cref{eq:metrics_theta}), and $\Xi'_x(\Omega)$ is the effective dephasing (\cref{eq:metrics_Xi'}).
The second portion of $S_{xx}^{\text{(quant)}}$ is the \emph{antisqueezing}, which is the main contribution of the antisqueezed quadrature of the field $\Tvec{s}$, which contributes to the test mass motion by an amount whose power spectral density is
  \begin{equation}
    \eta_x(\Omega) \left\{[1-\Xi'_x(\Omega)] \rme^{+2r} + \Xi'_x(\Omega) \rme^{-2r}\right\} \sin^2{\left[\phi + \theta_x(\Omega)\right]}.
    \label{eq:antisqueezing}
  \end{equation}
The effective dephasing contributes with a spectral density
     \begin{equation}
    \eta_x(\Omega) \Xi'_x(\Omega) \rme^{+2r} \cos^2{\left[\phi + \theta_x(\Omega)\right]}.
    \label{eq:total_dephasing}
  \end{equation}
The effective dephasing $\Xi_x'(\Omega)$ has two contributions which are budgeted separately: the first is the \emph{fundamental dephasing} $\Xi_x(\Omega)$ (\cref{eq:metrics_Xi}), and the second is the rms \emph{squeeze angle fluctuations} $\phi_{\text{rms}}$.
Following~\cite{McCuller:2021mbn}, these two contributions are combined as
\begin{equation}
  \Xi'_x(\Omega) = \Xi_x(\Omega) + \phi_{\text{rms}}^2 - 2\Xi_x(\Omega) \phi_{\text{rms}}^2.
  \label{eq:metrics_Xi'}
\end{equation}
Note that the motional dephasing $\Xi_x(\Omega)$ is predominantly due to the feedback damping and cannot be reduced by reducing optical losses, unlike measurement dephasing $\Xi_y(\Omega)$ which is due primarily to such losses.

Finally, losses coming from unsqueezed vacuum fields $\{\Tvec{a}_\mu\}$ entering the optical system at the locations shown in \cref{fig:signal_flow_full} are separately budgeted\,---\,in particular, losses entering the arm cavities, the signal extraction cavity, the filter cavity, the squeezed light injection path, and the optical readout path.

Any mismatch between the modes of the various optical cavities will also contribute to all of these noise sources in a potentially complicated frequency dependent manner depending on the exact character of the mismatch~\cite{McCuller:2021mbn}. We have investigated this and find that their contributions are unlikely to make a significant impact over the frequencies of interest and so omit these effects for simplicity.

In addition to quantum noise, the total budgeted noise in the test mass motion include a number of classical noises, which may be force noises, which directly drive the test masses (indicated by $F_{\text{ext}}$ in \cref{fig:signal_flow_reduced}), or sensing noises, which enter through the feedback control system used to trap and damp the oscillator (indicated by $x_{\text{sens}}$ in \cref{fig:signal_flow_reduced}).
The forces contributing to the overall force spectral density $S_{FF}$ are seismic noise, Newtonian noise, thermal noise from the suspensions supporting the test masses, and impacts from gas molecules striking the test masses; and the sensing noises contributing to $S_{xx}^{(\text{sens})}$ are coating and substrate thermal noise and gas-induced optical phase noise.
Note that in the usual language of optomechanics, optical shot noise is a sensing noise which is essentially treated as entering the system with $x_\text{sens}$. That is not the case here: this noise is caused by the various quantum vacua entering the apparatus, mostly the squeezed vacuum field $\Tvec{s}$ (as shown in \cref{fig:signal_flow_full}), and is accounted for in $S_{xx}^{(\text{quant})}$.\footnote{A sensing noise not considered here is the noise arising from the dark current of the photodetectors and their electronics. It is not accounted for in $S_{xx}^{(\text{quant})}$, but would enter the loop at the same node from which $y$ is derived, and thus contribute to the total noise with a different transfer function than the $\eff{X}$ used for the other sensing noises described here. The impact of this noise term is expected to be minimal.}
The classical noises included here comprise those noises that are typically considered ``fundamental'' in that they cannot be mitigated without a reworking of the detector's basic optical or mechanical characteristics.
In practice, additional classical noises of a technical nature (for example, electronics noise) can also drive the motion of the test masses, but we have not included these here.

\subsection{Choice of parameters}
\label{sec:parameter_choice}
There are several easily tunable detector parameters that can be used to manipulate the noise-induced motion of the differential test mass degree of freedom.
These are the trapping frequency $\eff{\Omega}$, the damping quality factor $\eff{Q}$, the injected squeeze amplitude $r$, and the injected squeeze angle $\phi$.
One could undertake a numerical optimization of these four parameters to find the minimum occupation number.
However, the noise factorization described in \cref{sec:budgeted_noises} enables a dimensional reduction of this search space.
Given a particular choice of $\eff{\Omega}$ and $\eff{Q}$, the squeezed state entering the interferometer's antisymmetric port arrives at the test mass with a frequency-dependent rotation $\theta_x(\Omega)$ (\cref{eq:metrics_theta}).
If $\theta_x$ does not vary strongly across the bandwidth of the oscillator, then it is advantageous to choose the injected squeeze angle to be $\phi \approx -\theta_x(\eff{\Omega})$, which will minimize the antisqueezing (\cref{eq:antisqueezing}).
Once this term is minimized, it remains to choose the squeeze amplitude $r$.
One notes that while increasing $r$ lowers the quantum noise from the injected squeezed vacuum (\cref{eq:injected_squeezing}), it makes the noise from total dephasing larger (\cref{eq:total_dephasing}); therefore, the optimal $r$ is found by balancing these two noises in the vicinity of $\eff{\Omega}$.
This leaves only the choice of $\eff{\Omega}$ and $\eff{Q}$.
Since the total motional spectrum contains various classical noises in addition to the quantum noise, we numerically searched over $\eff{\Omega}$ and $\eff{Q}$, optimizing $\phi$ and $r$ at each point, in order to find the combination of parameters that minimizes $\osc{\overline{n}}$.

We emphasize that this quantum noise injection strategy does not use a filter cavity, as is now standard practice in the LIGO detectors and is the baseline assumption for Cosmic Explorer and Einstein Telescope. The purpose of a filter cavity used for broadband sensitivity enhancement is to compensate for the rotation of the squeezed state at the antisymmetric port $\theta_y(\Omega)$ which is different from the rotation of the squeezed state responsible for the true motion of the mirrors $\theta_x(\Omega)$. In principle it would be possible to retune the filter cavity to compensate for $\theta_x(\Omega)$ instead, however we find that by making the frequency independent choice $\phi \approx -\theta_x(\eff{\Omega})$, the resulting anti-squeezing never makes a significant contribution to the total noise at the frequencies relevant for ground state cooling.

Finally, we note that some further optimization may be attainable by altering the homodyne angle $\zeta$ (\cref{eq:homodyne_angle}) or the detuning $\Delta$ of the signal extraction cavity~\cite{Mizuno:1993cj,Buonanno:2001cj}.
However, we did not find a strong dependence of the occupation number on these quantities.
We have therefore left $\Delta = 0$ and $\zeta = 0$ throughout. ($\zeta=0$ corresponds to the usual readout of the phase quadrature of the outgoing field.)
Varying either $\Delta$ or $\zeta$ will also alter the feedback control of the interferometer, and hence the dynamics of the trapped oscillator mode; sufficiently large $\Delta$ can provide an alternate route to trapping, in which case measurement-based feedback is used only to apply damping to the oscillator mode (as in Ref.~\cite{Corbitt:2007spn}).
In Fabry--Perot Michelson gravitational-wave interferometers, the sideband imbalance introduced by detuned signal extraction presents a significant control challenge, and was abandoned as an operating scheme for Advanced LIGO~\cite{Ward:2010qda}.
It is not planned as an operating scheme for Cosmic Explorer, but is part of the baseline for the low-frequency interferometer design in the Einstein Telescope, which also involves two filter cavities for operation as a gravitational-wave detector~\cite{ETDesign2020}.
Detuned operation can already be accommodated using the formalism presented in \cref{subsec:general_case}, and thereby extend the analysis of a three-mirror compound cavity in \cref{subsec:simplified_ifo}; we leave a systematic study of detuned operation for future work.%

\subsection{Results}
\label{sec:results}

\begin{figure*}[t]
  \centering
  \includegraphics[width=\textwidth]{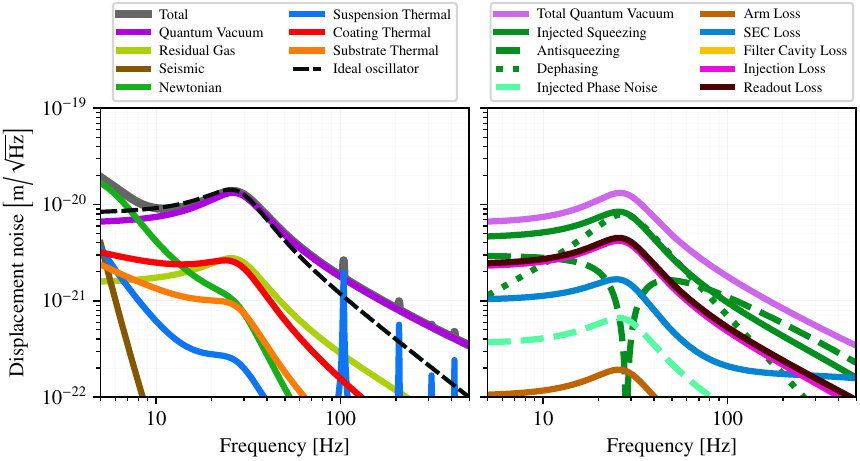}
  \caption{An example budget of the physical test mass motion for a trapped and cooled oscillator in a cryogenic silicon Cosmic Explorer, with frequency-independent squeezing.
  \label{fig:ce_physical_fis}
  }
\end{figure*}

We have computed budgets for the true motional spectrum $S_{xx}(\Omega)$ of a trapped and damped differential arm length degree of freedom for four gravitational-wave interferometers: LIGO A+, LIGO Voyager, Cosmic Explorer (CE), and CE Voyager.
In each of these four cases we have chosen the effective oscillator frequency $\eff{\Omega}$ to be in the region \qtyrange{20}{40}{\Hz}, finding that with a damping $\eff{Q} \lesssim 2$ it is possible to achieve small occupation numbers.
As described in \cref{sec:parameter_choice}, the squeeze parameters $\phi$ and $r$ are adjusted to minimize the spectrum of the motion in the vicinity of $\eff{\Omega}$, and hence minimize the occupation number $\osc{\overline{n}}$ as computed from \cref{eq:neff_quantum}.
For all four of the configurations we considered, occupation numbers at or near 1 are possible.

The relevant trapping and cooling parameters for the four detectors are given in \cref{tab:cooling_parameters}, along with the achieved occupation numbers.
For the two cases of LIGO A+ and CE Voyager, budgets are shown in \cref{fig:aplus_physical_fis,fig:ce_physical_fis}.
The left panels of these figures are budgets of the total noise, showing the importance of each of these noises, including a trace showing the total quantum noise.
Additionally, a dashed black line shows the corresponding thermal-noise limited spectrum of an ideal damped harmonic oscillator following \cref{eq:ideal_thermal}, with $\osc{T}$ determined by the relation $\osc{\overline{n}} = 1/(\rme^{\hbar\eff{\Omega}/k_{\text{B}} \osc{T}} - 1)$.
As anticipated, the true motion near and below \qty{10}{\Hz} is dominated by force noises, especially seismic noise and the thermal fluctuations of the test mass suspension.\footnote{We note that the total thermal noise of the suspension near \qty{10}{\Hz} is not, in general, simply the off-resonance thermal noise of the fundamental mechanical mode of the pendulum, but instead contains thermal fluctuations from other modes~\cite{Aston:2012ona,LIGO:2020xsf,Hall:2020dps}.}
In the case of Cosmic Explorer, the appearance of local gravity fluctuations also presents a force noise that drives the motion of the test masses (\cref{sec:newtonian_noise}).

\begin{figure}[t]
  \centering
  \includegraphics[width=\columnwidth]{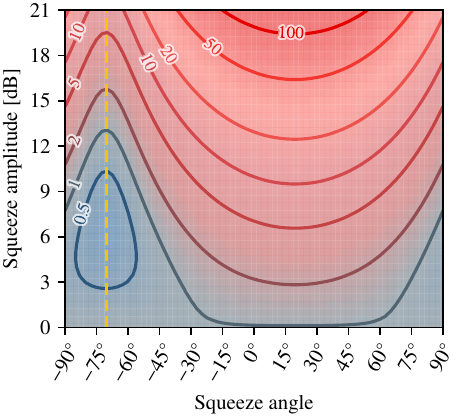}
  \caption{The occupation number as the injected squeeze angle $\phi$ and amplitude $r$ are varied for the configuration of LIGO A+ considered previously (\cref{fig:aplus_physical_fis,tab:cooling_parameters}).
    The dashed vertical line is $-\theta_x(\eff{\Omega})$, showing that the lowest occupation number is achieved when $\phi$ is tuned to counteract the rotation $\theta_x$ of the squeezed state.}
  \label{fig:aplus_sqz_opt}
\end{figure}

The right panel of each figure shows the individual noises contributing to the total quantum noise; these individual noises follow from the factorization described in \cref{subsubsec:general_spectra}.
In both cases one can see that the physically motivated factorization of the quantum noise clearly exhibits the desired tuning of the quantum noise, so that antisqueezing is minimized near $\eff{\Omega}$, and the contributions from injected squeezing and dephasing are balanced.
As an illustration of the success of this budget-based minimization procedure for the quantum noise, in \cref{fig:aplus_sqz_opt} we show the occupation number in LIGO A+ as injected squeeze angle $\phi$ and amplitude $r$ are varied.
Evidently, the configuration budgeted in \cref{fig:aplus_physical_fis} indeed achieves the lowest occupation number.

\begin{table*}[t]
  \begin{ruledtabular}
    \begin{tabular}{lSSSS}
 Parameter                                     &   {LIGO A+} &   {LIGO Voy.} & {CE}   & {CE Voy.}   \\
\hline
 Mass of test mass $M$ [kg]                              &    40   &   200    & 320    & 320        \\
 Arm power $P$ [MW]                                      &     0.8 &     3    & 1.5    & 12.0       \\
 Arm finesse $\Fa$ [---]                                 &   450   &  3100    & 450    & 1500       \\
 SEC finesse $\Fs$ [---]                                 &    18   &   140    & 310    & 1000       \\
 Oscillator frequency $\eff{\Omega}/2\pi$ [Hz]           &    37   &    32    & 23     & 28         \\
 SQL frequency $\sql/2\pi$ [Hz]                          &    65   &    37    & 7      & 15         \\
 $\Omega_x/2\pi$ [Hz]                                    &   117   &    41    & {---}  & {---}      \\
 Oscillator $\eff{Q}$ [---]                              &     1.2 &     1.4  & 1.8    & 1.5        \\
 Oscillator occupation $\overline{n}_{\text{osc}}$ [---] &     0.3 &     0.2  & 0.7    & 0.2        \\
 Oscillator occupation $\overline{T}_{\text{osc}}$ [nK]  &     1.3 &     1.0   & 1.3    & 0.9       \\
 Injected squeeze angle $\phi$ [deg.]                    &   -70   &   -49    & -4     & -11        \\
 Generated squeeze amplitude $r$ [dB]                    &     6   &     5    & 13     & 9          \\
 Squeezed field injection loss                           &     5\% &     5\%  & 3.0\%  & 3.0\%      \\
 Readout loss                                            &    10\% &     5\%  & 3.5\%  & 3.5\%      \\
 Homodyne angle $\zeta$ [deg.]                           &     0   &     0    & 0      & 0          \\
  \end{tabular}
  \end{ruledtabular}
  \caption{Relevant parameters for cooling current and future gravitational-wave detectors.}%
  \label{tab:cooling_parameters}
\end{table*}

%% file: sections/technical.tex
\section{Technical issues in cooling GW detectors}
\label{sec:technical}

Having budgeted the fundamental noises that limit the achievable cooling of the differential arm length degree of freedom in LIGO and Cosmic Explorer, we now turn to several technical issues that will require attention in cooling experiments with gravitational-wave detectors.

\subsection{Feedforward cancellation of other degrees of freedom}
\label{sec:feedforward}

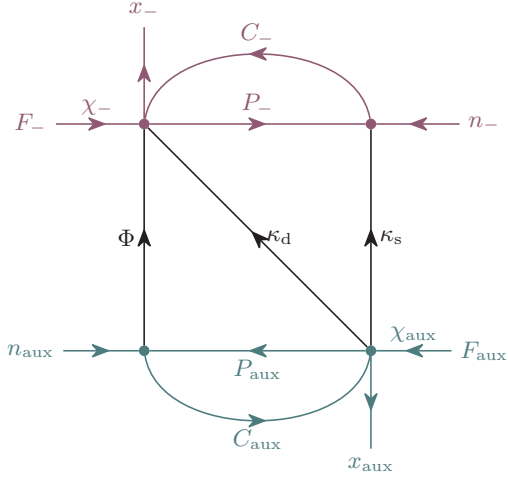
\begin{figure}
  \centering
    \begin{tikzpicture}[semithick]

    \begin{scope}[Magenta!50!Black]
      \node[dot] (Fd1) at (0,3) {};
      \node[dot] (nd1) at (3,3) {};
      \node (Fd) at (-1.5,3) {$F_-$};
      \node (xd) at (0,4.5) {$x_-$};
      \node (nd) at (4.5,3) {$n_-$};

      \draw[midarrow] (Fd) -- node[above] {$\chi_-$} (Fd1);      
      \draw[midarrow] (Fd1) -- node[above] {} (xd);
      \draw[midarrow] (Fd1) -- node[above] {$P_-$} (nd1);
      \draw[midarrow] (nd) -- node[above] {} (nd1);
      \draw[midarrow,bend right=80,looseness=1] (nd1) to node[above] {$C_-$} (Fd1);
    \end{scope}

    \begin{scope}[TealBlue!50!Black]
      \node (na) at (-1.5,0) {$\aux{n}$};
      \node[dot] (na1) at (0,0) {};
      \node (Fa) at (4.5,0) {$\aux{F}$};
      \node (xa) at (3,-1.5) {$\aux{x}$};

      \node[dot] (Fa1) at (3,0) {};
      \draw[midarrow] (Fa) -- node[above] {$\aux{\chi}$} (Fa1);
      \draw[midarrow] (na) -- (na1);
      \draw[midarrow] (Fa1) -- (xa);
      \draw[midarrow] (Fa1) -- node[below] {$\aux{P}$} (na1);
      \draw[midarrow,bend right=80,looseness=1] (na1) to node[below] {$\aux{C}$} (Fa1);
    \end{scope}

    \draw[midarrow] (Fa1) -- node[right] {$\kappa_{\text{d}}$} (Fd1);
    \draw[midarrow] (Fa1) -- node[right] {$\kappa_{\text{s}}$} (nd1);

    \draw[midarrow] (na1) -- node[left] {$\Phi$} (Fd1);

  \end{tikzpicture}
  \caption{Signal flow for the coupling of an auxiliary degree of freedom (labeled by ``aux'') to the differential arm length degree of freedom (labeled by ``$-$'').
    Each degree of freedom has an optical plant $P$ and a controller $C$, through which external force $F$ and sensing noise $n$ enter.
    The amount of true motion in each loop is labeled by $x$.
    The true motion $\aux{x}$ of the auxiliary degree of freedom can couple into the differential arm length either by appearing in the differential arm length sensor (indicated by the $\kappa_{\text{s}}$ path) or by directly imparting motion into the differential arm length (the $\kappa_{\text{d}}$ path).
     The measurement of the auxiliary degree of freedom is applied to the motion of the main degree of freedom through the feedforward filter $\Phi$ to partially cancel these couplings; however, the couplings of both $\aux{F}$ and $\aux{n}$ cannot be simultaneously subtracted since they enter the system at different locations.}
  \label{fig:aux_signal_flow}
\end{figure}

The measurement channel $y$ that provides the apparent motion of the differential arm length is sensitive to other degrees of freedom, particularly the Michelson length (differential length from the input test masses to the beamsplitter) and the length of the signal extraction cavity.
Displacements of these degrees of freedom can impart an additional sensing noise in the differential arm length readout, or it can cause true displacement of the differential arm length (e.g., by a radiation-pressure coupling)~\cite{Izumi:2016joj}.
In LIGO, these noises are fed forward to the differential arm length by actuating on the test masses, which reduces their appearance in the differential arm length readout, which is all that is required for maximal sensitivity to gravitational waves~\cite{Meadors:2013lja}.
However, to reduce the amount of \emph{true} differential arm length motion arising from these auxiliary degrees of freedom, as is desirable for cooling, the feedforward scheme would need to be altered to separately account for the couplings via sensing and couplings via displacement.

\cref{fig:aux_signal_flow} illustrates this feedforward.
True motion $\aux{x}$ of the auxiliary degree of freedom couples to the differential arm length degree of freedom, either by causing a true displacement in $x_-$ via the coupling $\kappa_{\text{d}}$, or by causing a sensing noise in $e_-$ via $\kappa_{\text{s}}$.%
\footnote{%
    Generically, one also can have coupling of the displacement $x_-$ into the auxiliary error signal $\aux{e}$ and auxiliary motion $\aux{x}$.
    Usually these couplings are ignored in the sensitive band of gravitational-wave detectors because
    the superior vibrational isolation and thermal noise performance of the test mass suspensions results in a motion $\aux{x}$ that is dominated by the motions of other optics in the optical path length of the auxiliary degree of freedom.
    For the auxiliary Michelson degree of freedom, this is the beamsplitter; for the auxiliary signal extraction degree of freedom, this is the signal extraction mirror and any relay mirrors between it and the beamsplitter.%
}
Since the true motion $\aux{x}$ is not, in general, experimentally observable, the auxiliary error signal $\aux{e}$ is used as a proxy to apply feedforward subtraction $\Phi$, which as drawn in \cref{fig:aux_signal_flow} is applied to the differential arm length loop directly as a displacement.

In the case of the Michelson degree of freedom, the $\kappa_{\text{d}}$ path is negligible and the coupling to the main differential arm degree of freedom is given by the frequency-domain relation
\begin{multline}
    \label{eq:mminus}
    x_- = G_- \chi_- F_- + H_- \frac{n_-}{P_-} \\ + \left(G_- \aux{G} \aux{P} \Phi + H_- \aux{G} \frac{\kappa_{\text{s}}}{P_-}\right) \aux{\chi} \aux{F} \\ + \left( G_- \aux{G} \aux{P} \Phi + H_- \aux{H} \frac{\kappa_{\text{s}}}{P_-}\right) \frac{\aux{n}}{\aux{P}},
\end{multline}
where $G = 1/(1 - CP)$ for both the auxiliary and differential arm degrees of freedom, and $H = CP G$.
If the coupling of auxiliary sensing noise $\aux{n}/\aux{P}$ is more significant than that of externally induced displacement $\aux{\chi}\aux{F}$, then $\Phi$ should be chosen to cancel that path: $G_- \aux{G} \aux{P} \Phi = -H_- \aux{H} \kappa_{\text{s}} / P_-$.%
\footnote{As noted above, in normal operation for gravitational-wave detection, the feedforward is tuned to minimize the appearance of the auxiliary noises in the error signal $e_-$, rather than the true motion $x_-$\,---\,see \cref{sec:ff_algebra}.}
The true differential motion $x_-$ then still contains a component $x_-^{\text{(ff)}}$ arising from the coupling of the displacement $\aux{\chi}\aux{F}$, given by%
\footnote{If the coupling of displacement noises is more significant than that of sensing noises, then $\Phi$ should rather be chosen to cancel the displacement noises, in which case the sensing noises of the auxiliary degree of freedom cause a true motion of the differential arm length
\begin{equation}
    x_-^{\text{(ff)}} = -H_- \frac{\kappas}{P_-} \frac{\aux{n}}{\aux{P}}.
\end{equation}}
\begin{equation}
    x_-^{\text{(ff)}} = H_- \frac{\kappa_{\text{s}}}{P_-} \aux{\chi} \aux{F}.
\end{equation}
If sufficiently large, this coupled motion could present an obstacle to cooling to the fundamental noise limits presented in \cref{sec:ligo_ce}.
The coupling factors $H_-$ and $\kappas/P_-$ cannot generally be freely varied: the form of the closed loop $H_-$ is set by the choice of trapping and damping, and in the vicinity of the trapping frequency is of order unity, and the magnitude of $\kappas/P_-$ is set by the detector's optical parameters (for the Michelson degree of freedom, $\kappas/P_-\simeq\pi/2\mathcal{F}_{\text{a}}$, where $\mathcal{F}_{\text{a}}$ is the arm finesse).
This means that the magnitude of $x_-^{\text{(ff)}}$ likely needs to be controlled by limiting the amount of auxiliary motion $\aux{x}$.
For Advanced LIGO, in which $\mathcal{F}_{\text{a}} \simeq 450$, externally driven motion $\aux{\chi}\aux{F}$ of the Michelson degree of freedom appears as a true differential arm length motion with a coupling factor of order 0.01.
We therefore expect that, with adequate feedforward, the Michelson length should not spoil the cooling of the differential arm length so long as the motion in this degree of freedom near the trapping frequency has amplitude spectral density of order $\qty{e-18}{\meter/\sqrt{\Hz}}$ or better.
If Cosmic Explorer chooses a similar arm cavity finesse, then the requirements on the Michelson motion are similar.
We have not attempted to estimate the requirements on the motion of the signal extraction cavity length, where the coupling depends on (among other things) static imbalances in length or alignment between the arms of the interferometer, which can be difficult to estimate from first principles.
This feedforward implementation needs to contend with displacement ($\kappa_{\text{d}}$) coupling in addition to sensing ($\kappa_{\text{s}}$) coupling.

\subsection{Local gravity fluctuations}
\label{sec:newtonian_noise}

Future ground-based gravitational-wave interferometers generally assume that the local gravity fluctuations perturbing the test masses will be subtracted from the gravitational-wave datastream using arrays of auxiliary sensors~\cite{Coughlin:2016tmu,Harms:2019dqi,Tringali:2019xqx,LIGO:2020xsf,Hall:2020dps,Badaracco:2023vpk}.
In the foregoing calculations we have not assumed any reduction of gravity fluctuations by subtraction, since the impact of any such subtraction on the true motion of the test masses depends on its implementation.
A realtime subtraction implemented by actuating on the test masses would appropriately reduce the true motion of the differential arm length degree of freedom, while an offline subtraction implemented in software would not.
Particularly for Cosmic Explorer, it is evident from \cref{fig:ce_physical_fis} that local gravity fluctuations add significant motion for trapping frequencies near and below \qty{10}{\Hz}; an appropriate realtime subtraction would therefore be advantageous if feedback cooling were desirable at these low frequencies.

\subsection{Feedback control performance}
\label{sec:feedback_performance}

Phase lags in the detector control system impose limitations on the trapping and damping of the oscillator.
Some of these phase lags arise from time delays, such as the light propagation time down the detector arms or the computational speed of the data acquisition and digital control system~\cite{Bork:2020zet}.
However, these delays overall are of order \qty{100}{\us} and therefore amount to a phase lag of only a few degrees for frequencies below \qty{100}{\Hz}~\cite{LIGOScientific:2017aaj}.
Other phase lags arise from the controller; in particular, the feedback control in Advanced LIGO must be attenuated above a few hundred hertz, both to conserve the finite bandwidth of the test mass actuator and to avoid exciting mechanical resonances in the test mass suspensions.
This attenuation induces phase lag that increases with frequency, thereby setting an upper limit on the achievable trapping frequency (in \textcite{Whittle:2021mtt}, this limit was about \qty{150}{\Hz}).
A quantitative analysis of the controller phase delay in Cosmic Explorer is premature, but given these considerations a trap frequency above \qty{100}{\Hz} may not be attainable.

\subsection{Quantum readout challenges}
\label{subsec:new_technology_challenges}

The lowest occupation numbers in \cref{tab:cooling_parameters} are achieved in the cryogenic silicon realizations of LIGO and Cosmic Explorer.
Shifting to cryogenic silicon as a gravitational-wave detector technology requires a number of innovations described in detail elsewhere (e.g., Refs.~\cite{LIGO:2020xsf,Hall:2020dps,PostO5}).
Among other things, these realizations require a laser wavelength longer than the \qty{1064}{\nm} currently employed by LIGO, Virgo, and Kagra, and here we comment on the feasibility of squeezed vacuum generation and detection at longer wavelengths.
At \qty{1550}{\nm}, the generation and detection of squeezed states in excess of \qty{12}{\dB} is already possible, using cavity-enhanced optical parametric amplifiers (OPAs) with quadratically nonlinear materials, similar to the strategy at \qty{1064}{\nm}~\cite{Mehmet:2011je,Meylahn:2022qvj}.
However, in the \qty{2}{\um} region, which may be selected for cryogenic silicon interferometry and is the assumption in our models, work is ongoing and may require further departures from the current squeezed light implementations.

Multiple tabletop experiments with OPAs using quadratic nonlinearity in the \qty{2}{\um} region have demonstrated squeezed light production; an experiment at \qty{1984}{\nm} inferred the generation of \qty{11}{\dB} of squeezed light~\cite{Mansell:2018tbn,Yap:2019gqt}, and a broadly similar level of squeezed light generation can be inferred from an experiment conducted at \qty{2128}{\nm}~\cite{Darsow-Fromm:2021zif}.
This level of generated squeezing is sufficient to achieve the ground state cooling schemes presented here; the cryogenic silicon Cosmic Explorer scheme, which is more stringent in this respect than the LIGO Voyager scheme, requires \qty{9}{\dB} of squeezing to be generated and then propagated to the dark port with less than \qty{3}{\%} injection loss.
Also, the level of squeezed light generation required for cooling test mass motion is generally less than the target level anticipated for gravitational wave detection (\qty{18}{\dB} for LIGO Voyager and both realizations of Cosmic Explorer).

An unmet challenge with \qty{2}{\um} interferometry is overcoming readout loss due to the subunity quantum efficiency of available photodiodes; this loss (indicated as $\epsilon_\text{ro}$ in \cref{fig:signal_flow_full}) superposes unsqueezed vacuum on the optical state just before homodyne detection, hence adding quantum noise to the measurement.
The extended indium--gallium--arsenide photodiodes used in Ref.~\cite{Darsow-Fromm:2021zif} contributed a readout loss of \qty{8(3)}{\%}, whereas the total readout loss required in the cryogenic LIGO and Cosmic Explorer configurations is \qty{5}{\%} and \qty{3.5}{\%}, respectively.
This total readout loss requirement jointly constrains the subunity quantum efficiency in photodetection and also the optical loss from mirrors, lenses, Faraday isolators, and spatial modecleaning cavities in the readout chain; these readout chain losses may amount to ${}\lesssim\qty{2}{\%}$~\cite{PostO5}, meaning that a quantum efficiency of \qty{99}{\%} or better should be targeted to meet the overall readout loss requirement.
If this quantum efficiency target cannot be reached, it may be circumvented by the inclusion of optical amplification before homodyne detection~\cite{Frascella:2020zuo}.
A recent experiment used two OPAs, one to generate a squeezed vacuum state at \qty{2}{\um} and a second to amplify the state before detection on a homodyne apparatus, to improve the effective readout loss from \qty{26}{\%} (limited by the photodiode quantum efficiency) to \qty{5}{\%}~\cite{Kwan:2025yyz}.
Sum frequency generation with a shorter-wavelength pump field may offer another route to overcoming readout loss by transferring the \qty{2}{\um} output field to a wavelength where high quantum efficiency photodiodes are available~\cite{2014PhRvL.112g3602V,VoyagerDesign,BolingbrokeThesis}.
Additionally, optomechanical implementations of output amplification have been proposed~\cite{Bai:2019gyd}.

For all configurations considered in \cref{tab:cooling_parameters}, including the room-temperature fused silica interferometers, achieving the required readout loss also requires mitigating loss from the aggregate effects of mismatched transverse optical modes, and nonzero reflectivities at intervening optical interfaces (e.g., from antireflection coatings) in the readout chain.
This is required not only for cooling experiments, but for sensitive operation as gravitational-wave detectors.

%% file: sections/conclusion.tex
\section{Conclusion}
\label{sec:conclusion}

The main goal of this work has been to extend the standard methods for computing quantum noise in optomechanical systems to include the effects of feedback control and to clarify the propagation of force and sensing noises throughout such systems.
Though our formalism is general, we have exercised it here by applying it to the differential test mass motion of gravitational-wave detectors, operated with measurement-based feedback control that synthesizes a trapped and cold-damped oscillator.
We have exhibited a physically motivated decomposition of the dominant quantum noises in the interferometer, giving some insight into how to appropriately tune the optomechanical configuration to minimize these quantum noises for such an oscillator.
With budgets of the fundamental noise sources\,---\,both quantum and classical\,---\,contributing to the total motion of the cold-damped oscillator, we estimate the effective occupation number of the oscillator by computing its noise power, though we note that the utility of such a computation is limited for occupation numbers significantly below 1 in the bandlimited regime heretofore considered for gravitational-wave detectors.
We have also pointed out several technical issues that need to be considered in cooling experiments in cooling gravitational-wave detectors.

There are several extensions to our work.
First, our model of feedback control has assumed a filter that applies a proportional and derivative feedback only.
This creates the dynamics appropriate to synthesize a viscously damped oscillator at the desired trapping frequency, but if some deviation from such dynamics is allowable, perhaps off resonance from the effective oscillator frequency, it is possible that a different feedback filter may provide superior performance\,---\,e.g., by reducing the amount of control applied above the oscillator resonance, thereby injecting less noise from the measurement.
One way to tailor the feedback algorithmically is to recast the problem in terms of optimal control, using %
the measurement $y$ to construct a state estimator $\hat{x}$ whose statistics (particularly its variance) could be computed, serving as a proxy for the unobservable true motion $x$.
In the case of a viscously damped oscillator probed by (and subject to the radiation pressure of) a single optical field, the state space construction of the system with feedback is relatively straightforward, and one can use the tools of linear quadratic control to devise an optimal feedback scheme in the sense of minimizing a cost function quadratic in the state vector, similar to the classical case~\cite{Doherty:1998zln}.
The construction of a state-space model for the system considered here is more complicated, since it involves several (in principle, arbitrarily many) quantum noise fields, as well as frequency-domain relations among the fields, the mechanical degree of freedom, and the measurement.
However, it is possible in principle to construct a state space representation if the state vector is enlarged to track the evolution of the quantum noise fields in addition to the mechanical degree of freedom; in this procedure, the state vector likely also needs to be augmented to capture the frequency-domain relations between the physical variables, and the spectral shape of the noise terms entering the system.
We note that optimal control has already been explored in cavity optomechanical systems~\cite{Wieczorek:2015fba,Miki:2022sff}, including levitated systems~\cite{Conangla:2018nnn,Magrini:2020agy}, and there are a number of theoretical proposals exploiting it in the context of gravitational entanglement experiments (see below).

A second set of extensions concerns optomechanical systems with elements that we have not explicitly considered here.
In particular, we have not provided expressions for measurement-based feedback with nonzero cavity detuning, although our formalism in \cref{subsec:general_case} allows for it; this would be useful, for instance, in an analysis of sideband cooling.
We have also pointed out proposals or experiments that employ optical amplification before the measurement, especially in the context of overcoming optical loss in photodetection.
Such apparatus could be included in our formalism as well so long as they admit a two-photon description of their operation on optical fields, as all of the possibilities discussed in \cref{subsec:new_technology_challenges} do.
Extension of our formalism to other systems, such as optically levitated systems, is also likely possible by introducing appropriate two-photon loss fields to account for laser recoil heating.

Finally, we comment on the application of the scheme described here to fundamental physical tests with gravitational-wave detectors and other optical interferometers.
The relation between mechanical motion and its measurement in a quantum-noise-dominated regime is central to proposals seeking to witness entanglement of gravitational-wave interferometer test masses~\cite{Mueller-Ebhardt:2007ehs,Mueller-Ebhardt:2009bfo}, or to exploit the entanglement of low-occupation-number test masses with the optical field to study, e.g., decoherence~\cite{Marshall:2002exi}.
Characterization of test mass entanglement is also essential for proposals for interferometers with suspended or freefalling masses placed in close gravitational contact with one another (e.g., Refs.~\cite{Schnabel:2015ada,Miao:2019pxw,Datta:2021ywm,Miki:2024qcz,Matsumoto:2025pdt}).
Regardless of whether the entanglement performance is expressed in terms of phonon occupation number (cf. Ref.~\cite{Direkci:2023bme}, for instance), it is key to have a model that accurately factors in all quantum and classical noise contributions, and distinguishes between the apparent motion inferred from the measurement and the true motion of the system.

%% file: sections/radiation_pressure_mirror.tex
\section{Radiation pressure dynamics of a mirror}
\label{sec:mirror_radiation_pressure_dynamics}

\input{sections/signal_flow_mirror}

In this appendix we explicitly reduce the signal flow graph for the radiation pressure dynamics of a single mirror illuminated with a laser incident on the front surface, and we show that it reproduces the well-known expressions found elsewhere~\cite{2012emqm.book.....M,Danilishin:2012fa}.\footnote{This process is equivalent to performing Gaussian elimination on the matrix relating the nodes to each other.}
The signal flow graphs are shown in \cref{fig:signal_flow_mirror}. The first graph reproduces the end test mass subset of the graph of the full interferometer \cref{fig:signal_flow_full}. If the laser light has power $P$ and wavenumber $k$, then the carrier-frequency ($\Omega = 0$) electric fields entering and exiting the mirror surfaces are
\begin{equation}
  \Tvec{E}_\text{fi} = \sqrt{P}\,\Tvec{e}_q,\ 
  \Tvec{E}_\text{fo} = -r\Tvec{E}_\text{fi},\ 
  \Tvec{E}_\text{bo} = t\Tvec{E}_\text{fi},\ 
  \Tvec{E}_\text{bi} = 0.
  \label{eq:dc_fields}
\end{equation}
We are not interested in the fields exiting the back surface of the mirror and so will not keep track of signals flowing to $\Tvec{a}_\text{bo}$.

The vectors that map the mirror motion to the excitation of optical fields in the phase quadrature are
\begin{equation}
  \Tvec{z} \equiv \Tvec{z}_\text{f}
  = 2kr\Tmat{R}(\pi/2) \Tvec{E}_\text{fi} = 2kr\sqrt{P}\,\Tvec{e}_p.
\end{equation}
The radiation pressure from each optical node $n$ is $\pm 2\Tvec{E}_n^\dag/c$, where $c$ is the speed of light, $+$ is for the radiation pressure sourced from the two front nodes, and $-$ is for the radiation pressure sourced from the two back nodes. Therefore, the radiation pressure forces produced from each optical node are described by dotting the node into one of the dual vectors
\begin{equation}
  \Tvec{f}^\dag \equiv \Tvec{f}_\text{fi}^\dag
  = \frac{2}{c}\sqrt{P}\,\Tvec{e}_q^\dag,\ 
  \Tvec{f}_\text{fo}^\dag = -r\Tvec{f}^\dag,\ 
  \Tvec{f}_\text{bo}^\dag = -t\Tvec{f}^\dag,\ 
  \Tvec{f}_\text{bi}^\dag = 0.
\end{equation}

With these expressions, several nodes can be eliminated from the first graph in \cref{fig:signal_flow_mirror} by composing the operators along each path. The path going from $\Tvec{a}_\text{fi}$ to $\Tvec{a}_\text{bo}$ and then to the node probed by $x$ is $t\chi_0\Tvec{f}_\text{bo}$. This path is added to the path going from $\Tvec{a}_\text{fi}$ directly to the node probed by $x$ to give the path shown.
Note that the path $\Tvec{z}_\text{b} \propto \Tvec{e}_p$ can be discarded because the only subsequent path leaving the node $\Tvec{a}_\text{bo}$ is $\Tvec{f}^\dag_\text{bo} \propto \Tvec{e}^\dag_q$, and the product $\Tvec{f}_\text{bo}^\dag \Tvec{z}_\text{b}$ is always zero\,---\,i.e., the mirror motion only impresses phase sidebands onto the field leaving the back side of the mirror, and these sidebands cannot produce a radiation force to react back on the mirror motion.

Next, the node probed by $x$ is eliminated to go from the second to the third graph. Again two paths are added to produce the path from $\Tvec{a}_\text{fi}$ to $\Tvec{a}_\text{fo}$. In all cases care is taken to compose operators backward along a path to maintain the ordering of the signal flow.

Finally, the radiation pressure loop is eliminated from the third diagram to arrive at the final diagram. The general rule is that when a loop with open loop propagator $\Tmat{F}$ is eliminated, its node is replaced with the corresponding closed-loop propagator $\Tmat{G} = (\Tmat{1} - \Tmat{F})^{-1}$. Using the fact that
\begin{equation}
  \left(\Tmat{1} + a\Tvec{e}_p \Tvec{e}_q^\dag\right)^{-1}
  = \Tmat{1} - a\Tvec{e}_p \Tvec{e}_q^\dag
\end{equation}
for any scalar $a$, the closed-loop propagator for the radiation pressure loop is
\begin{equation}
  \Tmat{G}_\text{rp}
  = \left(\Tmat{1} + r\chi_0\Tvec{z}\Tvec{f}^\dag\right)^{-1}
  = \Tmat{1} - r\chi_0\Tvec{z}\Tvec{f}^\dag.
\end{equation}
Noting that $\Tvec{f}^\dag \Tvec{z}\propto \Tvec{e}_q^\dag \Tvec{e}_p = 0$, the reflection from the mirror\,---\,i.e., the path from $\Tvec{a}_\text{fi}$ to $\Tvec{a}_\text{fo}$\,---\,is therefore
\begin{align}
  \Tmat{H}
  &= \Tmat{G}_\text{rp} \left[-r\Tmat{1} + \chi_0(1 - t^2)\Tvec{z}\Tvec{f}^\dag\right] \nonumber\\
  &= -r\Tmat{1} + \chi_0 \left(1 - t^2 + r^2\right) \Tvec{z}\Tvec{f}^\dag \nonumber\\
  &= -r\left[\Tmat{1} - \frac{4k(2\mathcal{R} + \mathcal{L})\chi_0 P}{c} \Tvec{e}_p\Tvec{e}_q^\dag \right] \nonumber\\
  &= -r\left(\Tmat{1} - \Krp \Tvec{e}_p\Tvec{e}_q^\dag\right)
  = -r \begin{bmatrix}
    1 & 0 \\
    -\Krp & 1
    \end{bmatrix},
\end{align}
where the optomechanical coupling is
\begin{equation}
  \Krp = \frac{4k(2\mathcal{R} + \mathcal{L})\chi_0 P}{c}
  \label{eq:general_rp_coupling}
\end{equation}
and we used $r^2 + t^2 = 1 - \mathcal{L}$.
Next, the transmission from $\Tvec{a}_\text{fi}$ to $x$ is the sum of two paths:
\begin{align}
  \Tvec{D}^\dag &= \chi_0 (1 - t^2)\Tvec{f}^\dag - r\chi_0\Tvec{f}^\dag\Tmat{H} \nonumber\\
  &= (2\mathcal{R} + \mathcal{L})\chi_0\Tvec{f}^\dag \nonumber\\
  &= \frac{2(2\mathcal{R} + \mathcal{L})\chi_0\sqrt{P}}{c}\,\Tvec{e}_q^\dag \nonumber\\
  &= \frac{2(2\mathcal{R} + \mathcal{L})\chi_0\sqrt{P}}{c}
  \begin{bmatrix}
    1 & 0
  \end{bmatrix}.
  \label{eq:mirror_force_to_motion}
\end{align}
The conversion of forces $F_\text{ext}$ to reflected field $\Tvec{a}_\text{fo}$ is
\begin{equation}
  \Tvec{Z} = \Tmat{G}_\text{rp} \chi_0\Tvec{z} = \chi_0\Tvec{z}
  = 2kr\chi_0\sqrt{P}\,\Tvec{e}_p
  = 2kr\chi_0\sqrt{P}
  \begin{bmatrix}
    0 \\
    1
  \end{bmatrix}.
  \label{eq:mirror_motion_to_field}
\end{equation}
Finally, the effective optomechanical susceptibility is the sum of two paths; however, the one going through the radiation pressure loop is zero in this case since it is proportional to $\Tvec{e}_q^\dag \Tvec{e}_p$:
\begin{equation}
  \chi_\text{om} = \chi_0 - r\chi_0^2\Tvec{f}^\dag\Tmat{G}_\text{rp}\Tvec{z} = \chi_0.
\end{equation}

%% file: sections/signal_flow_mirror.tex
\begin{figure*}[t]
  \centering
  \begin{tikzpicture}[semithick]
    \begin{scope}[shift={(-5,0)}]
      \node[draw,thick,rounded corners,align=center,gray] at (-0.25,4.75) {Step 1};
      
      \node[dot] (etm_fo) at (0,0) {};
      \node[dot] (etm_bi) at (4,0) {};
      \node[dot] (etm_fi) at (0,4) {};
      \node[dot] (etm_bo) at (4,4) {};

      \node (a_fi) at (-1,4) {$\Tvec{a}_{\text{fi}}$};
      \node (a_fo) at (-1,0) {$\Tvec{a}_{\text{fo}}$};
      \node (a_bi) at (5,0) {$\Tvec{a}_{\text{bi}}$};
      \node (a_bo) at (5,4) {$\Tvec{a}_{\text{bo}}$};

      \draw[midarrow] (a_fi) to (etm_fi);
      \draw[midarrow] (etm_fo) to (a_fo);
      \draw[midarrow] (a_bi) to (etm_bi);
      \draw[midarrow] (etm_bo) to (a_bo);

      \node[dot] (x1) at (2,2) {};
      \node[dot] (f_fi) at (1.15,3.15) {};
      \node[dot] (f_fo) at (1.15,0.85) {};
      \node[dot] (f_bi) at (2.85,0.85) {};
      \node[dot] (f_bo) at (2.85,3.15) {};

      \draw[midarrow] (etm_bi) -- node[below right] {$r\Tmat{1}$} (etm_bo);
      \draw[midarrow] (etm_fi) -- node[above left] {$-r\Tmat{1}$} (etm_fo);
      \draw[midarrow,bend left=15] (etm_fi) to node[above] {$t\Tmat{1}$} (etm_bo);
      \draw[midarrow,bend left=15] (etm_bi) to node[below right] {$t\Tmat{1}$} (etm_fo);

      \draw[midarrow,bend right=20] (x1) to node[below right] {$\Tvec{z}_{\text{b}}$} (etm_bo);
      \draw[midarrow,bend right=20] (x1) to node[above left] {$\Tvec{z}_{\text{f}}$} (etm_fo);

      \draw[midarrow] (etm_fi) to node[below] {$\Tvec{f}_{\text{fi}}^\dagger$} (f_fi);
      \draw[midarrow] (etm_fo) to node[below right] {$\Tvec{f}_{\text{fo}}^\dagger$} (f_fo);
      \draw[midarrow] (etm_bi) to node[above] {$\Tvec{f}_{\text{bi}}^\dagger$} (f_bi);
      \draw[midarrow] (etm_bo) to node[above left] {$\Tvec{f}_{\text{bo}}^\dagger$} (f_bo);
      \draw[midarrow] (f_fi) to node[below left] {$\chi_0$} (x1);
      \draw[midarrow] (f_fo) to node[below right] {$\chi_0$} (x1);
      \draw[midarrow] (f_bi) to node[above right] {$\chi_0$} (x1);
      \draw[midarrow] (f_bo) to node[above left] {$\chi_0$} (x1);

      \node (x) at (5,2.75) {$x$};
      \node (Fext) at (1.5,-1) {$F_{\text{ext}}$};
      \node[inner sep=0] (x_sens) at (0,-1) {$x_{\text{sens}}$};

      \draw[midarrow] (x_sens) -- node[left] {$\Tvec{z}_{\text{f}}$} (etm_fo);
      \draw[midarrow,bend right=10] (x1) to (x);
      \draw[midarrow,bend right=20] (Fext) to (f_fo);
    \end{scope}

    \begin{scope}[shift={(3.5,0)}]
      \node[draw,thick,rounded corners,align=center,gray] at (-0.25,4.75) {Step 2};
      \node[dot] (etm_fo) at (0,0) {};
      \node[dot] (etm_fi) at (0,4) {};

      \node (a_fi) at (-1,4) {$\Tvec{a}_{\text{fi}}$};
      \node (a_fo) at (-1,0) {$\Tvec{a}_{\text{fo}}$};

      \draw[midarrow] (a_fi) to (etm_fi);
      \draw[midarrow] (etm_fo) to (a_fo);

      \node[dot] (x1) at (2.25,2) {};

      \draw[midarrow] (etm_fi) to node[above left] {$-r\Tmat{1}$} (etm_fo);

      \draw[midarrow,bend left=30] (x1) to node[below right] {$\Tvec{z}$} (etm_fo);

      \draw[midarrow] (etm_fi) to node[above right] {$\chi_0(1 - t^2)\Tvec{f}^\dag$} (x1);
      \draw[midarrow] (etm_fo) to node[above left] {$-r\chi_0\Tvec{f}^\dag$} (x1);

      \node (x) at (5,2.75) {$x$};
      \node (Fext) at (4.25,-0.5) {$F_{\text{ext}}$};
      \node[inner sep=0] (x_sens) at (0,-1) {$x_{\text{sens}}$};

      \draw[midarrow] (x_sens) to node[left] {$\Tvec{z}$} (etm_fo);
      \draw[midarrow,bend right=10] (x1) to (x);
      \draw[midarrow,bend left=10] (Fext) to node[below left] {$\chi_0$} (x1);
    \end{scope}

    \begin{scope}[shift={(-5,-7)}]
      \node[draw,thick,rounded corners,align=center,gray] at (-0.25,4.75) {Step 3};
      \node[dot] (etm_fo) at (0,0) {};
      \node[dot] (etm_fi) at (0,4) {};
      \node[dot] (x) at (4,4) {};
      \node[dot] (Fext) at (4,0) {};

      \node (a_fi) at (-1,4) {$\Tvec{a}_{\text{fi}}$};
      \node (a_fo) at (-1,0) {$\Tvec{a}_{\text{fo}}$};
      \node (x_tp) at (5,4) {$x$};
      \node (Fext_exc) at (5,0) {$F_{\text{ext}}$};

      \draw[midarrow] (a_fi) to (etm_fi);
      \draw[midarrow] (etm_fo) to (a_fo);
      \draw[midarrow] (x) to (x_tp);
      \draw[midarrow] (Fext_exc) to (Fext);

      \draw[midarrow] (etm_fi) to node[above] {$\chi_0(1 - t^2)\Tvec{f}^\dag$} (x);
      \draw[midarrow] (Fext) to node[right] {$\chi_0$} (x);
      \draw[midarrow] (Fext) to node[below] {$\chi_0 \Tvec{z}$} (etm_fo);
      \draw[midarrow] (etm_fi) to node[above right] {$-r\Tmat{1} + \chi_0(1 - t^2)\Tvec{z}\Tvec{f}^\dag$} (etm_fo);
      \draw[midarrow,bend right=30] (etm_fo) to node[right] {$-r\chi_0\Tvec{f}^\dag$} (x);
      \draw[midarrow,looseness=80] (etm_fo) to[out=15,in=75] node[above] {$-r\chi_0\Tvec{z}\Tvec{f}^\dag$} (etm_fo);

      \node[inner sep=0] (x_sens) at (0,-1) {$x_{\text{sens}}$};
      \draw[midarrow] (x_sens) to node[left] {$\Tvec{z}$} (etm_fo);
    \end{scope}

    \begin{scope}[shift={(3.5,-7)}]
      \node[draw,thick,rounded corners,align=center,gray] at (-0.25,4.75) {Step 4};
      \node[dot] (etm_fo) at (0,0) {};
      \node[dot] (etm_fi) at (0,4) {};
      \node[dot] (x) at (4,4) {};
      \node[dot] (Fext) at (4,0) {};

      \node (a_fi) at (-1,4) {$\Tvec{a}_{\text{fi}}$};
      \node (a_fo) at (-1,0) {$\Tvec{a}_{\text{fo}}$};
      \node (x_tp) at (5,4) {$x$};
      \node (Fext_exc) at (5,0) {$F_{\text{ext}}$};

      \draw[midarrow] (a_fi) to (etm_fi);
      \draw[midarrow] (etm_fo) to (a_fo);
      \draw[midarrow] (x) to (x_tp);
      \draw[midarrow] (Fext_exc) to (Fext);

      \draw[midarrow] (etm_fi) to node[above] {$\Tvec{D}^\dag = (2\mathcal{R} + \mathcal{L})\chi_0\Tvec{f}^\dag$} (x);
      \draw[midarrow] (Fext) to node[right] {$\chi_0$} (x);
      \draw[midarrow] (Fext) to node[below] {$\Tvec{Z}=\chi_0\Tvec{z}$} (etm_fo);
      \draw[midarrow] (etm_fi) to node[above right] {$\Tmat{H} = -r\left(\Tmat{1} - \Krp\Tvec{e}_p\Tvec{e}_q^\dag\right)$} (etm_fo);

      \node[inner sep=0] (x_sens) at (0,-1) {$x_{\text{sens}}$};
      \draw[midarrow] (x_sens) to node[right] {$\Tvec{Y}=\Tvec{z}$} (etm_fo);
    \end{scope}

  \end{tikzpicture}
  \caption{Signal flow graph for the radiation pressure dynamics of a single mirror illuminated with a laser from the front surface.
    The full graph in Step 1 is reduced to the graph seen in Step 4.
    See \cref{sec:mirror_radiation_pressure_dynamics} for details.}
  \label{fig:signal_flow_mirror}
\end{figure*}
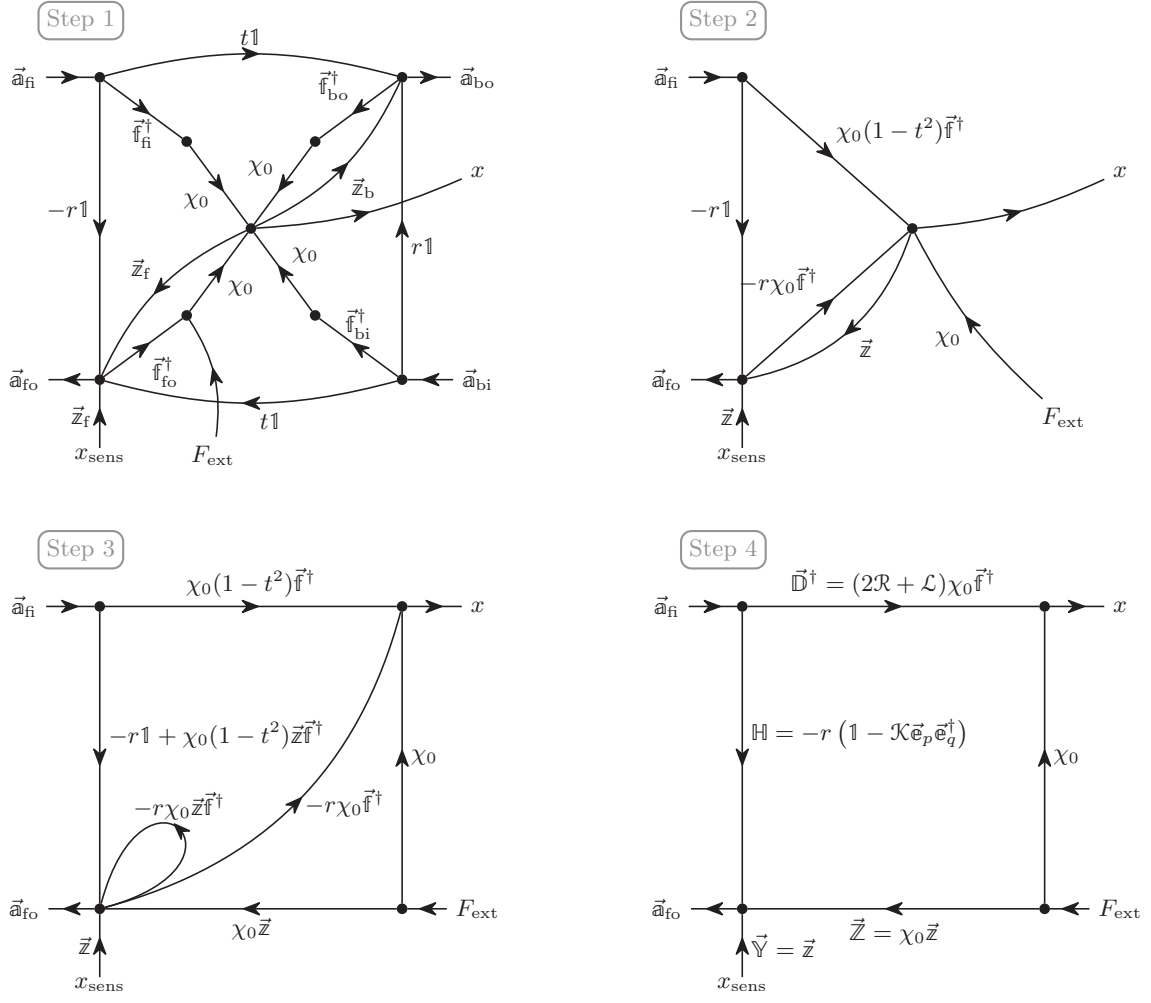

%% file: sections/ff_app.tex
\section{Feedforward loop algebra}
\label{sec:ff_algebra}

As described in \cref{sec:feedforward}, feedforward cancellation is usually employed to cancel the appearance of noises induced by the auxiliary degrees of freedom in the differential arm length error signal $e_-$.
On the other hand, in cooling experiments we rather want to minimize the appearance of these noises in the true motion $x_-$.

Both $\aux{x}$ and the displacement-referred error signal $\aux{e}/\aux{P}$ contain contributions from external force $\aux{F}$ and sensing noise $\aux{n}$, but loop algebra (using the graph in \cref{fig:aux_signal_flow}) shows these two terms appear differently:
\begin{align}
    \aux{x} &= \aux{G} \aux{\chi}\aux{F} + \aux{H} \frac{\aux{n}}{\aux{P}} \\
    \frac{\aux{e}}{\aux{P}} &= \aux{G} \aux{\chi}\aux{F} + \aux{G} \frac{\aux{n}}{\aux{P}}
\end{align}
The resulting true motion in the differential arm length is
\begin{align}
  x_- &= G_- \chi_- F_- + H_- \frac{n_-}{P_-} \nonumber \\
      &\hphantom{==} {} + G_- \left[\Phi \aux{e} + (\kappad + C_- \kappas) \aux{x}\right], \\
\end{align}
while the displacement-referred error signal is
\begin{align}
  \frac{e_-}{P_-} &= x_- + \frac{n_-}{P_-} + \frac{\kappas}{P_-} \aux{x} \nonumber \\
      &= G_- \chi_- F_- + G_- \frac{n_-}{P_-} \nonumber \\
      &\hphantom{==} {} + G_- \left[\Phi \aux{e} + \left(\kappa_{\text{d}} + \frac{\kappas}{P_-}\right) \aux{x}\right].
\end{align}

With regard to the coupling via sensing $\kappas$, we note that auxiliary motion $\aux{x}$ appears in the true differential motion $x_-$ with a factor $G_- C_- \kappas$, while it appears in the error signal $e_- / P_-$ with a factor $G_- \kappas / P_-$, which differs from the former factor by a ratio $G_- / (G_- - 1)$.
Thus, in any experiment where the feedforward $\Phi$ has been carefully tuned to suppress the appearance of $\aux{x}$ in the differential arm length error signal $e_- / P_-$, this suppression will not carry over to the true differential motion $x_-$.